# Nationally Scalable Hydrogen Fueling Infrastructure Deployment: A Megaregion Analysis and Optimization Approach

Vivek Sujan, Junchaun Fan, Gurneesh Jatana, Ruixiao Sun

Oak Ridge National Laboratory

## Abstract

Decarbonizing regional and long-haul freight is challenging due to the limitations of battery-electric commercial vehicles and infrastructure constraints. Hydrogen fuel cell medium- and heavy-duty vehicles (MHDVs) offer a viable alternative, aligning with the decarbonization goals of the Department of Energy and commercial entities. Historically, alternative fuels like compressed natural gas and liquefied propane gas have faced slow adoption due to barriers like infrastructure availability. To avoid similar issues, effective planning and deploying zero-emission hydrogen fueling infrastructure is crucial. This research develops deployment plans for affordable, accessible, and sustainable hydrogen refueling stations, supporting stakeholders in the decarbonized commercial vehicle freight system. It aims to benefit underserved and rural energy-stressed communities by improving air quality, reducing noise pollution, and enhancing energy resiliency. This research also provides a blueprint for replacing diesel in over-the-road Class 8 freight truck applications with hydrogen fueling solutions. The study focuses on the Texas Triangle Megaregion (I-45, I-35, and I-10), the I-10 corridor between San Antonio, TX, and Los Angeles, CA, and the I-5/CA-99 corridors between Los Angeles, CA, and San Francisco, CA. This area represents a significant portion of U.S. heavy-duty freight movement, carrying ~8.5% of the national freight volume. Using the OR-AGENT (Optimal Regional Architecture Generation for Electrified National Transport) modeling framework, the study conducts an advanced assessment of commercial vehicles, road and freight networks, and energy systems. The framework integrates data on freight mobility, traffic, weather, and energy pathways to deliver a region-specific, optimized vehicles powertrain architectures, infrastructure deployment solutions, operational logistics, and energy pathways. By considering all vehicle origin-destination pairs utilizing these corridors and all feasible fueling station location options, the framework's genetic algorithm identifies the minimum number and optimal locations of hydrogen refueling stations, ensuring no vehicle is stranded. It also determines fuel schedules and quantities at each station. A roadmap for station deployment based on multiple adoption trajectories ensures a strategic rollout of hydrogen refueling infrastructure.

## Introduction

The need for commercial vehicle decarbonization through Zero Emission Vehicle (ZEV) technology is critical, as the transportation sector remains the largest energy consumer in the United States, accounting for approximately 37% of all energy used in 2023—see Figure 1 [1]. When electricity-related emissions are distributed to economic end-use sectors, transportation activities contributed to 28.5% of U.S. greenhouse gas emissions in 2022 [2]. The largest sources of transportation greenhouse gas emissions in 2022 were light-duty trucks, which include sport utility vehicles, pickup trucks, and minivans, accounting for 36.5% of emissions. MHDVs were responsible for 22.9% of emissions, followed by passenger cars at 20.4%.

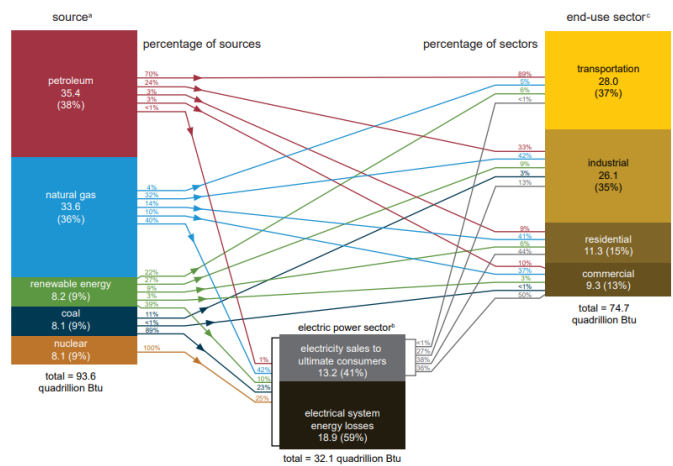

*Figure 1. 2023 U.S. energy consumption by source and sector [1]*

Other significant contributors to transportation-related emissions included commercial aircraft at 7.2%, pipelines at 3.8%, ships and boats at 2.8%, other aircraft at 2.0%, and rail at 2.0%. These figures reflect direct emissions of $CO_2$, $CH_4$, and $N_2O$ from fossil fuel combustion, as well as indirect emissions from electricity use and emissions from non-energy sources, such as lubricants. Additionally, hydrofluorocarbon emissions from mobile air conditioners and refrigerated transport are factored into the calculations for these vehicle types—see Figure 2 and Figure 3. By transitioning to ZEV technologies, such as battery-electric and hydrogen fuel cell vehicles, the commercial transportation sector can reduce its carbon footprint, enhance energy efficiency, and better integrate with renewable energy sources. This shift is essential for aligning with national and global decarbonization goals and fostering a cleaner, more sustainable transportation future [3].

The heavy reliance on fossil fuels, especially in medium- and heavy-duty vehicles, drives significant greenhouse gas emissions, intensifies climate change, and worsens air quality, particularly in densely populated urban areas [3,4]. Transitioning to ZEV technologies, such as battery-electric and hydrogen fuel cell vehicles, offers a promising solution by reducing the carbon footprint of the commercial transportation sector. This shift enhances energy efficiency and enables better integration with renewable energy sources. Ultimately, adopting ZEV technology is essential for meeting national and global decarbonization targets, ensuring a cleaner environment, and paving the way for a sustainable future in transportation [3].

Vehicle electrification, encompassing both battery-electric and hydrogen fuel cell technologies, is a vital element of transportation decarbonization strategies for MHDVs [5]. Achieving the scale of electric vehicle deployment necessary for substantial emissions reductions requires forward-thinking planning to ensure that clean fueling infrastructure is widely available and operated efficiently to maintain low costs and support broader cross-sectoral decarbonization goals [5]. This is especially crucial for MHDV freight operations, as these vehicles require reliable and consistent access to charging and hydrogen refueling stations across multiple jurisdictions to meet their fueling needs [5]. However, the development of this infrastructure will likely be led by private sector stakeholders, and the energy demands of MHDV will be significant, necessitating coordination across regions and among various stakeholders. These vehicles typically operate along key freight corridors that link inland and ocean ports, freight hubs, and industrial clients. Therefore, establishing a clear, integrated planning process for deploying charging and refueling infrastructure along these corridors, at freight depots, and at port facilities is essential. This will ensure that the network meets the needs of all MHDV users, minimizes impacts on the power grid, and aligns with environmental equity and justice (EEJ) goals [5]. Such comprehensive planning will also provide confidence to infrastructure developers and fleet operators, streamlining the process and speeding up deployment. By having these plans in place, the corridors will be ready for the rapid rollout of high-power charging infrastructure through larger deployment programs supported by federal, state, or private funding sources.

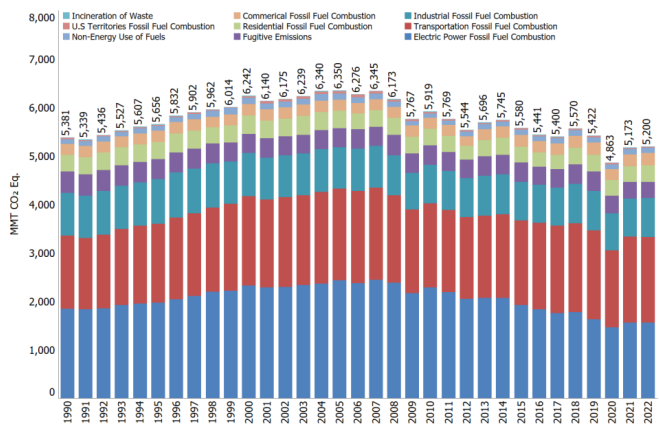

*Figure 2. Trends in energy sector GHG emissions [2].*

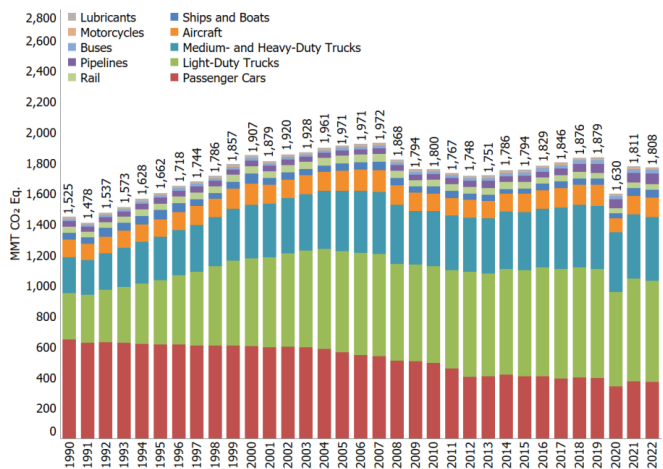

*Figure 3. Trends in transportation-related GHG emissions [2]*

In this research, we create detailed plans for commercial vehicle electrification and refueling/recharging infrastructure across several key areas: the Texas Triangle Megaregion, the I-10 corridor extending west from Houston, TX to the Port of Los Angeles/Long Beach in California, and the route between Los Angeles and San Francisco along the I-5 and CA-99 corridors. This corridor group is known as the Houston to Los Angeles (H2LA) corridors for this project. These regions are crucial for U.S. freight movement, encompassing major transportation routes within Texas and California, and linking two of the country's largest ports—see Figure 4 [6]. The Houston area is expected to emerge as a major hub for clean hydrogen production, acting as a vital source of hydrogen fuel for this infrastructure network [7]. This initiative aims to support the transition to clean energy by enhancing hydrogen refueling capabilities in key freight corridors, thereby facilitating more sustainable transportation across these significant routes.

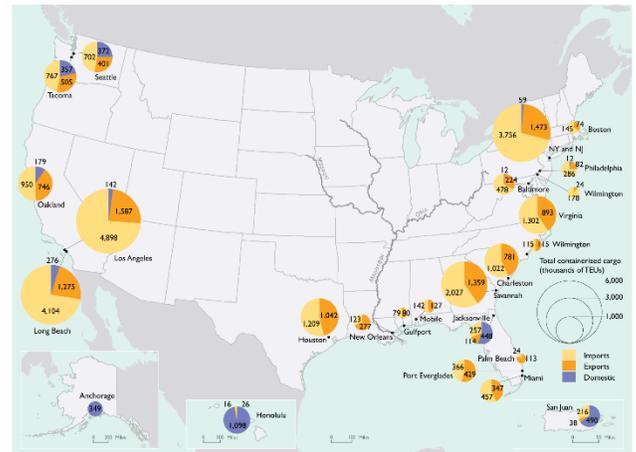

*Figure 4. Top 25 water ports by TEU [6]*

Megaregions are large, interconnected geographic areas that encompass multiple metropolitan areas and their surrounding regions [8]. They are characterized by significant economic, social, and infrastructure linkages that extend beyond individual cities or metropolitan areas. Megaregions often feature:

- **Economic Integration**: High levels of economic activity and interdependence among cities and surrounding areas, often driven by shared industries, labor markets, and supply chains.

- **Transportation Networks**: Well-developed and extensive transportation infrastructure that connects various cities and regions, facilitating the movement of people and goods.

- **Shared Resources and Challenges**: Common environmental, social, and planning issues, such as air quality, water resources, and land use, that affect multiple areas within the megaregion.

- **Urban Growth**: Significant urban and suburban development that often leads to a blending of metropolitan areas into a larger, more cohesive economic and social unit.

These megaregions are important for understanding regional planning, economic development, and infrastructure needs on a larger scale than individual cities or metropolitan areas. The Texas Triangle Megaregion includes five of the 20 largest cities in the U.S. and is home to over 70% of the state's population—nearly 21 million people—see Figure 5. This region is anchored by Texas' four major urban centers: Austin, Dallas–Fort Worth, Houston, and San Antonio, which are connected by Interstate 45, Interstate 10, and Interstate 35. As one of the nation's eleven megaregions, the Texas Triangle plays a vital role in freight movement, handling 306 million ton-miles of daily truck freight, which represent 5.3% of the total U.S. truck freight activity, facilitated by approximately 35,700 miles of daily commercial vehicle miles traveled (VMT). Additionally, the I-10 freight corridor from San Antonio to Los Angeles supports another 118 million ton-miles of daily freight movement, contributing 2.1% of the total U.S. truck freight volume—see Figure 6 [9]. This combined activity underscores the Texas Triangle and the I-10 corridor as critical arteries in the national supply chain, facilitating the flow of goods between Texas, California, and the broader U.S. economy. As such, this region is a focal point for infrastructure development, including efforts to expand clean fuel technologies like hydrogen to support sustainable freight and logistics systems.

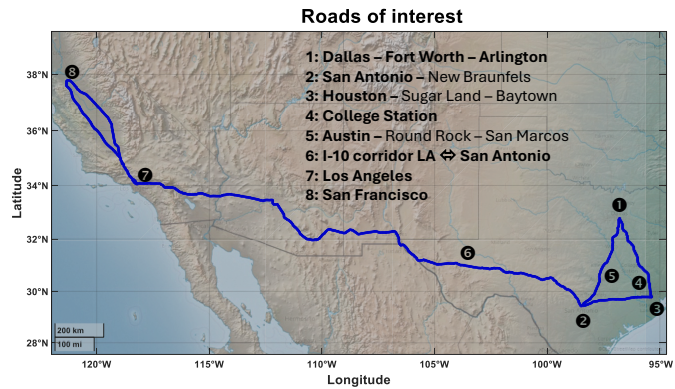

Figure 5. Texas Triangle Megaregion, I-10 corridor, and I-5/CA-99 corridors of interest

| Year | Type | Truck Kton-miles (Annual) | | | | |
|---|---|---|---|---|---|---|
| | | (A) Texas Triangle | (B) I10 corridor | A + B | US | % Share of A + B |
| 2017 | Single | 28,533,150 | 9,083,618 | 37,616,768 | 560,873,845 | 6.71% |
| 2017 | Combination | 83,122,180 | 33,942,284 | 117,064,464 | 1,538,794,017 | 7.61% |
| 2017 | Total | 111,655,327 | 43,025,903 | 154,681,230 | 2,099,667,813 | 7.37% |
| | | Truck VMT (Daily) | | | | |
| | | (A) Texas Triangle | (B) I10 corridor | A + B | US | % Share of A + B |
| | | 4,175,820 | 1,634,757 | 5,810,576 | 88,740,317 | 6.55% |
| | | 8,853,270 | 3,790,342 | 12,643,612 | 172,422,852 | 7.33% |
| | | 13,029,089 | 5,425,101 | 18,454,190 | 261,163,164 | 7.07% |

Figure 6. Summary of freight movement statistics in Texas Triangle and I-10 region [9]

In March 2024, the National Zero-Emission Freight Corridor Strategy was unveiled by the Joint Office of Energy and Transportation, alongside the U.S. Department of Energy (DOE), in collaboration with the Department of Transportation (DOT) and the Environmental Protection Agency (EPA) [5]. This strategy outlines the roadmap for deploying zero-emission charging and hydrogen fueling infrastructure for medium- and heavy-duty vehicles from 2024 through 2040. It aims to address the rising market demand by strategically directing public investments to support private sector advancements, streamline utility and energy regulatory planning, align industrial efforts, and improve air quality in communities heavily affected by diesel emissions. The H2LA corridor, covering about 25% of the national freight hubs identified in the first two phases of the strategy, is a key focus. Alongside this initiative is the Regional Clean Hydrogen Hubs program, or H2Hubs, launched in September 2022. With up to $7 billion allocated, the program aims to establish seven regional clean hydrogen hubs across the U.S. Seven hubs were selected for award negotiations in October 2023 [10]. These hubs will connect hydrogen producers, users, and infrastructure to boost hydrogen adoption. A key goal of the H2Hubs program is to match the growth in clean hydrogen production with increasing regional demand, thereby creating large-scale, commercially viable hydrogen ecosystems. The hubs will demonstrate the viability of low-carbon, hydrogen-based energy systems that can replace current carbon-intensive methods. The H2LA corridor, connecting the California and Gulf Coast Hydrogen hubs, is positioned to interlink directly with other major hubs, including those in the Pacific Northwest, Midwest, and Appalachia via interstate routes—see Figure 7. This makes it a pivotal part of the national hydrogen hub network, enabling the balancing of supply and demand across regions.

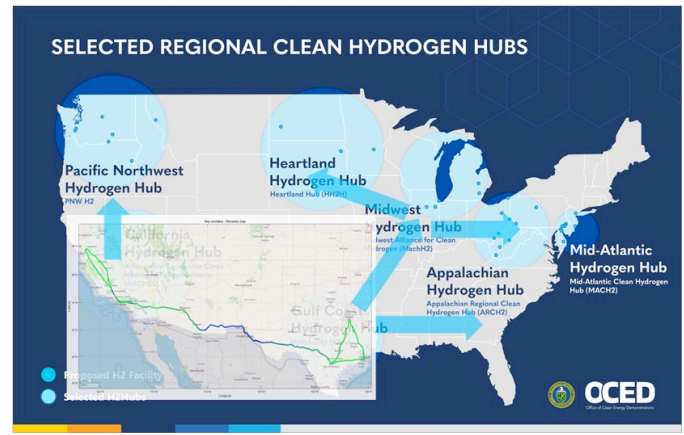

Figure 7. Selected regional clear hydrogen hubs [10] and the H2LA corridor network

In addition to the Clean Hydrogen Hubs program, The National Electric Vehicle Infrastructure (NEVI) Program is a federal initiative created under the Infrastructure Investment and Jobs Act (IIJA) to accelerate the development of BEV charging infrastructure across the U.S. [11,12]. The program's primary aim is to establish a nationwide network of BEV charging stations that will support the growing adoption of electric vehicles, particularly along interstate highways and major travel corridors. This initiative is a key component of the U.S. government's efforts to promote cleaner transportation and reduce greenhouse gas emissions from the transportation sector. The NEVI program has allocated $5 billion in funding over a five-year period (2022-2026) to help states and regions build and expand their BEV charging infrastructure. A significant portion of this funding is directed toward the development of fast-charging stations along designated Alternative Fuel Corridors, ensuring that drivers have reliable access to charging stations during long-distance travel. In addition to building the infrastructure, the program mandates that stations meet specific technical standards, such as minimum charging speeds, consistent reliability, and user-friendly payment systems. Furthermore, the NEVI program places a strong emphasis on equity and accessibility, ensuring that charging infrastructure is fairly distributed across the country, including underserved and rural areas. It encourages public-private partnerships, urging states to collaborate with private sector companies and utilities for the implementation and maintenance of charging stations. By establishing this robust national network, the NEVI program plays a critical role in advancing the U.S.'s climate goals, supporting the transition to a sustainable electric vehicle fleet, and promoting widespread BEV adoption.

In this paper, we introduce an innovative method for siting alternative fuel stations specifically tailored for commercial vehicles, addressing the unique challenges posed by hydrogen fueling stations and electric vehicle charging infrastructure. This method is designed to align with the infrastructure demands of major national programs, such as the National Zero-Emission Freight Corridor Strategy and the NEVI program. By strategically optimizing the placement of fueling and charging stations, the approach ensures that medium- and heavy-duty vehicles have efficient and reliable access to clean energy along key freight corridors, thus facilitating uninterrupted freight movement and connecting critical regions. This method is essential in the broader effort to advance clean energy infrastructure and supports the widespread deployment of zero-emission vehicles, ultimately contributing to national and global decarbonization objectives. It has been successfully applied to the H2LA corridor, a key freight route, and is readily scalable to other corridors and regions worldwide. The strength of this method lies in its use of detailed multi-agent vehicle modeling within real-world, complex environments. Through a systematic optimization process, supported by high-resolution data, the method provides accurate geo-located solutions for siting energy replenishment facilities along key transport routes. This approach not only ensures operational efficiency but also maximizes the economic and environmental benefits of zero-emission vehicle adoption.

This paper is structured as follows. In the next section, we present the overall methodology developed for this research, detailing the approach and framework used in the siting process. Following that, we explore the results and analysis, examining the outcomes of the study and their implications. We then focus on key attributes of the solutions, discussing the differences between Fuel Cell Electric Vehicles (FCEVs) and hydrogen refueling infrastructure, compared to BEVs and electric charging infrastructure. Finally, we conclude with remarks summarizing the findings and offering insights for future work.

## Methodology

Oak Ridge National Laboratory (ORNL) has developed an advanced commercial vehicle road freight network and energy systems architecture analysis framework called the OR-AGENT (Optimal Regional Architecture Generation for Electrified National Transport) modeling framework. This system-of-systems analytics platform has been introduced in publications [13,14] and developed extensively [15] for the workflow illustrated in Figure 8. Within the OR-AGENT framework, parametric optimization is employed to design vehicle powertrain architectures, local energy dispensing infrastructure (such as refueling and recharging systems), and regional energy infrastructure (including the electric grid, distributed energy resources (DER), and grid-scale storage). This optimization process addresses multiple stakeholder objectives, such as minimizing levelized system costs, total cost of ownership (TCO), carbon emissions, and may be readily adapted for others. The framework integrates data and models across various subsystems, covering electrified vehicle powertrain architectures, freight logistics, traffic patterns, road conditions, weather impacts, and energy flow pathways. By merging insights from vehicle operations, logistics, and energy infrastructure, OR-AGENT generates region-specific, seasonally optimized, and constraint-aware solutions for both vehicle and infrastructure deployment. The optimization process is driven by techno-economic metrics tailored to the needs of stakeholders, including fleet operators, equipment suppliers, energy providers, utilities, and local planning agencies. Employing a cost function that reflects these diverse interests, OR-AGENT serves as a versatile planning tool, enabling stakeholders such as governments, industries, and energy suppliers to strategically deploy electrified freight systems.

One of the primary strengths of the OR-AGENT framework is its ability to account for regional variations and navigate the complex, often competing objectives of multiple stakeholders. Unlike traditional approaches, which tend to handle infrastructure planning in isolation, OR-AGENT adopts a holistic perspective. It provides a coherent roadmap for the simultaneous development of vehicle and energy infrastructures. While this study focuses on heavy-duty trucks and the infrastructure needed for hydrogen and electric refueling, OR-AGENT's flexibility extends to a broader spectrum of vehicle energy solutions. It can evaluate powertrain configurations, energy sources, and infrastructure requirements for diesel, natural gas, hydrogen, and electricity across both on- and off-road vehicles. This versatility makes OR-AGENT a vital tool in designing sustainable, future-proof transport and energy systems across diverse regions and use cases. Although this paper applies a subset of OR-AGENT's capabilities, as seen in Figure 8, a summary of the overall workflow follows.

To summarize, the key outputs of this model include:

- Vehicle powertrain architecture recommendations
- Local infrastructure architecture (type, location, quantities of chargers/ refueling systems)
- Regional infrastructure architecture (grid and DER asset deployment)

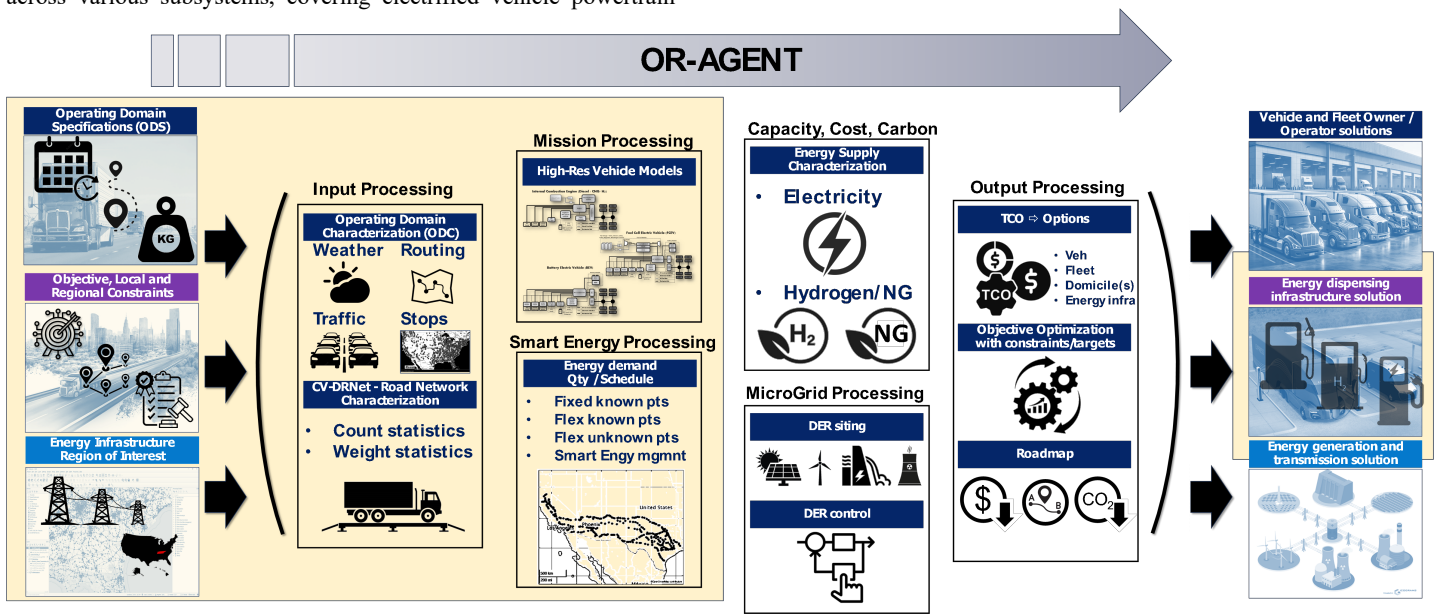

*Figure 8. OR-AGENT workflow construct (highlighted region is the focus of this research paper)*

Rather than delve into the details of each subsystem illustrated in Figure 8, we provide only a brief overview, as the majority of these details have already been extensively covered in previous publications or submissions and will be highlighted as such. In addition, for the scope of this paper not all subsystems have been exercised and will be indicated as such in the subsequent descriptions.

A. **Input requirements**:

- **Operating Domain Specifications (ODS):** Every study on an interconnected vehicle system begins by defining the specific customer use case being investigated. In this study, the focus is on heavy-duty freight trucks operating within the H2LA corridors. Key parameters include the vehicle type (class/weight), origin-destination (OD) coordinates (if available in GPS form), and the departure/operating schedules of the trucks. To avoid complications arising from time zone differences, all vehicle movements are modeled in Coordinated Universal Time (UTC), with the option to convert results back to local time if requested by stakeholders. For this study, the vehicle type is restricted to Class 8 combination tractor-trailer units, representing the primary freight movers in this corridor. Since we are analyzing a broad region of freight movement along the H2LA corridor, specific customer OD inputs are not directly available. Instead, these are derived as part of the *Operating Domain Characterization (ODC) Input processing*, which will be discussed later. Truck departure schedules are aligned with publicly available data on heavy-duty truck movements from ports, as illustrated in Figure 9 [16]. Future iterations will refine these inputs by incorporating more region-specific information based on higher-resolution data, as outlined in the ODC Input processing step.

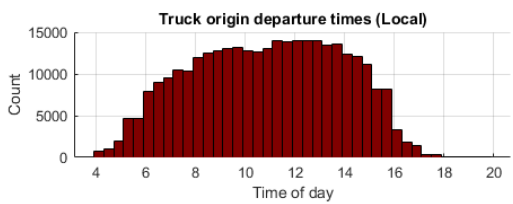

*Figure 9. Assumed truck departure times from origins*

- **Objective Function with Local and Regional Constraints**: In the context of this study, the objective function serves as the mathematical representation of the primary stakeholder goal, which is to optimize key attributes of the output such as the TCO, optimize the domicile infrastructure, minimize the effective system carbon emissions, or some other formulation. The optimization is subject to local and regional constraints. Local constraints include factors like grid capacity at specific station locations, land availability, fleet behind the fence infrastructure limitations or public access infrastructure siting constraints, and proximity to major freight routes or industrial hubs. These factors affect how many stations can be built, their size, and the available energy or hydrogen supply. On a regional level, constraints encompass broader factors like regulatory policies (ex. Advanced Clean Fleet), regional energy supply limitations, environmental impacts, and coordination across state boundaries. Constraints can either be limits or exclusions applied to a parametric search space of solutions, or targets that must be achieved. By balancing these local and regional constraints within the objective function, the model aims to provide a feasible, efficient infrastructure network that aligns with both immediate operational needs and long-term regional goals for decarbonization and sustainable freight transport. For the H2LA corridors the goal is to **minimize the total number of stations** while meeting vehicle demand. The optimization of this will be discussed in the **Output Processing step**.

- **Energy Infrastructure in the Region of Interest**: A key step in planning alternative fuel stations, especially hydrogen fueling and electric vehicle charging stations, is understanding the region where these stations will be located. This requires assessing the energy infrastructure in the area and how it connects to broader networks, such as roads, electric grids, and pipelines. Defining these boundaries can be challenging since regions are part of larger, interconnected systems that often extend beyond state and national borders. The complexity arises because energy infrastructure—like power grids and hydrogen production networks—operates across systems that need to be considered when placing stations. For example, a BEV charging station depends on local grid capacity but may also rely on power plants far outside the region. Similarly, hydrogen fueling stations are linked to production facilities through pipelines or long-distance transport. Regions are shaped by overlapping systems that must be coordinated. Road networks for commercial vehicles often span multiple utility areas, each with its own regulations and limits. Local factors such as economic activity, population, and freight movement further complicate the task of setting clear boundaries without overextending them. To manage this complexity, stakeholders should integrate data from transportation models, grid assessments, and logistics to define practical boundaries. A multi-layered approach is needed—one that considers both local infrastructure and broader networks linking the region to neighboring areas. This ensures a more resilient solution, supporting the smooth operation of alternative fuel vehicles and aligning with national decarbonization goals. For the H2LA corridor, Freight Analysis Framework (FAF) and National Household Travel Survey (NHTS) data [9,17] were used to identify the region of interest, shown in Figure 10, where the impact of commercial freight diminishes significantly outside of this area. This approach ensures focused infrastructure development without expanding the region unnecessarily.

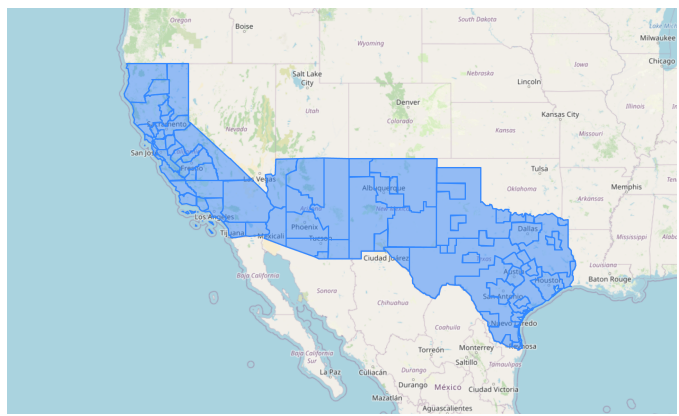

*Figure 10. H2LA region of interest approximated from FAF/NHTS. Sub-boundaries indicate NHTS relevant zones (583 zones nationally and 76 along the H2LA corridor)*

B. **Input Processing – Operating Domain Characterization**

- **Weather Processing:** Ambient conditions such as air temperature and pressure can significantly influence vehicle energy consumption by affecting various components and driving loads. These factors can alter the energy required for thermal management systems, accessory loads, and even change air density, impacting aerodynamic drag on the vehicle. Previous research has shown that seasonal variations alone can cause up to a 20% difference in energy consumption on the same route [18]. In this H2LA corridor study, weather-related factors such as temperature, air density, and wind speed/direction are incorporated using data from the National Oceanic and Atmospheric Administration (NOAA), alongside vehicle counts and schedules (departure/arrival times) for port-related traffic along each route. This data, developed through prior work [19], feeds into a comprehensive dataset integrated into an automated process, streamlining the establishment of ODS for specific ports or fleet domiciles. The factors used to develop the ODS are regionally and temporally specific, ensuring that the model captures the local variations that affect vehicle performance, making it adaptable for future studies.

- **Routing:** When OD information is provided, including GPS coordinates and schedules, direct vehicle routing can be established using tools such as the Google Directions API for general vehicle navigation or PC*Miler[1] for truck-specific directions. These GPS coordinates are mapped to elevation and road speed limits through the HERE Technologies Map Attributes API[2]. To calculate the directional slope, a MATLAB® acausal filter function, *filtfilt* is used along with a max slope clamp based on US highway engineering limits [20]. To efficiently handle large sets of OD pairs, this process has been automated within MATLAB, where HERE Technologies' Road Elevation data is stored in a comprehensive database, built through systematic screening of Level 1 to 4 roads (functional classification) across the U.S.

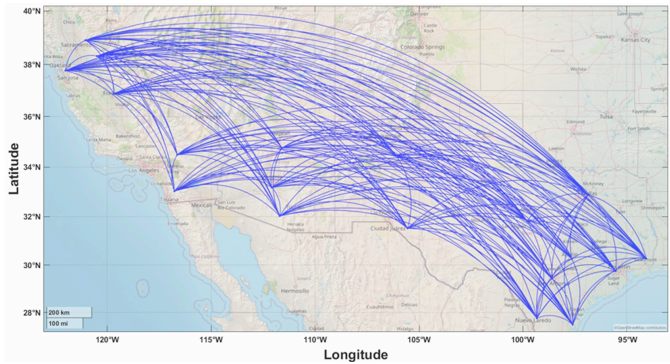

*Figure 11. H2LA ODs approximated from FAF/NHTS*

In cases where OD information is not initially provided, several other data sources can be used to generate this information. For instance, StreetLight Data[3] is a commonly used provider of highly granular vehicle OD information, broken down by Census Block Zones, covering about 217,526 zones nationwide. This dataset provides detailed insights into vehicle type, schedules, and other movement attributes. Alternatively, the OR-AGENT framework can generate a statistical view of OD patterns based on FAF and NHTS data, providing insights on OD distributions by calendar quarter and by FAF zones, as detailed in [21]. As indicated above, for the H2LA corridor in this paper, FAF and NHTS data are used to determine the statistical OD in the region of interest. See Figure 11, Figure 12, and Figure 13 for the OD and routing. The process of integrating higher resolution StreetLight data is underway.

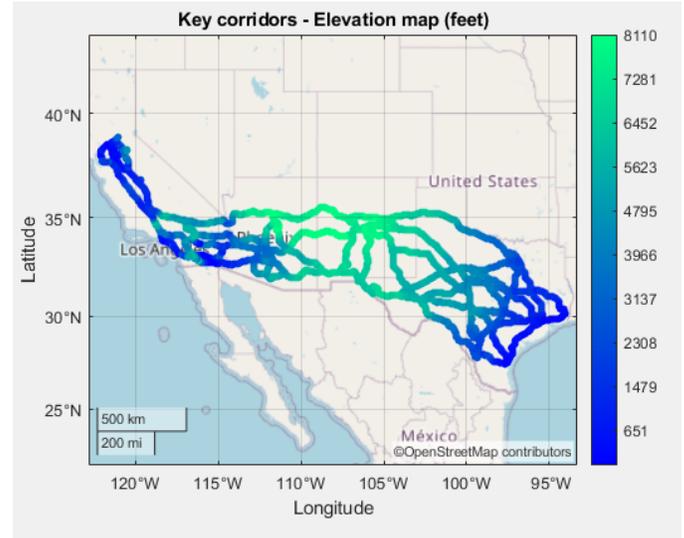

*Figure 12. Routes and elevations with additional roads of interest*

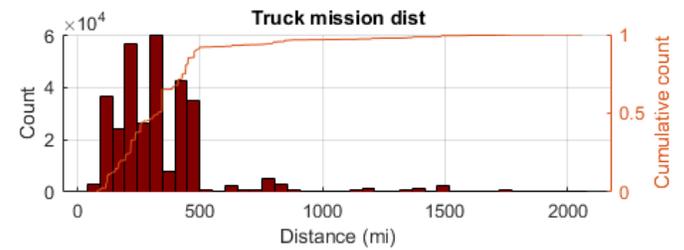

*Figure 13. OD route distance statistics*

- **Admissible Stops:** Admissible known or unknown stops are developed next. If the list of stops has been provided (known) as part of the constraint space (for example all current public access heavy-duty refueling stations or all private vehicle domiciles within a fleet network may be candidates for future zero-emission vehicle recharging or refueling points), then stations within some constraint may be defined as candidates. For the H2LA corridor the stops are the current public access heavy duty truck diesel refueling stations with d = 5 miles—impact shown in Figure 14.

If the list of stops is not provided but rather need to be discovered, then a list of candidate stop locations is generated by simply dividing up the region into a raster scan of candidate sites defined by a specific geometric construct For example, the region may be subdivided into an array of sub regions each measuring a predefined dLatitude x dLongitude or $dL_1(m)$ x $dL_2(m)$ These become the candidate locations for the refueling/recharging stations, and the reduction process described above may now be applied.

- **Traffic**: Accurately modeling naturalistic driving behavior hinges on effectively capturing traffic patterns. The approach being developed leverages HERE Technologies' traffic analytics, which provide average vehicle speed data in 15-minute intervals over the course of a week for any specific location (https://www.here.com/developer). While this data may not be highly granular, it offers a practical means of synthesizing realistic driving conditions based on real-world observations. By incorporating these traffic dynamics into the model, the system can better reflect the variability in vehicle speeds due to congestion, road conditions, and time-of-day effects, which is crucial for modeling energy consumption and operational schedules. Though still in development, this traffic integration within the OR-AGENT framework has the potential to significantly enhance the accuracy of naturalistic driving simulations. As such this feature is not yet incorporated into the H2LA corridor analysis. Further refinements and results from this work will be detailed in future iterations of this research. Other data source for similar but higher granularity data is being explored, including RITIS (https://ritis.org/login?r=Lw==).

---

[1]PC*Miler, https://www.truckingoffice.com/lp/pc-miler/
[2] HERE Technologies Map Attributes API, https://www.here.com/developer/
[3]StreetLight Data, https://www.streetlightdata.com/

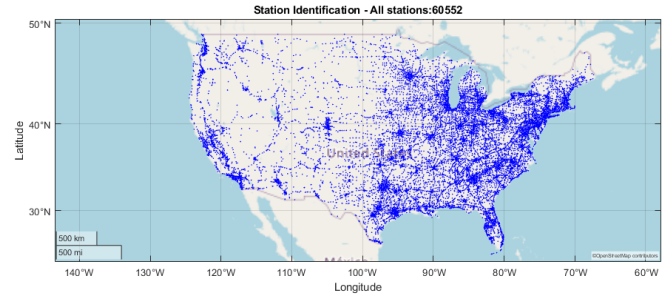

(a) All Heavy-Duty truck diesel refueling stations (Geotab-2021 https://www.gpsfms.com/)

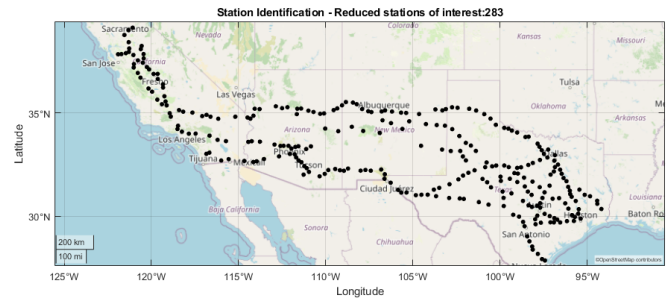

(b) Reduced candidate stations within 5 miles of the H2LA corridor and broader roads of interest

Figure 14. H2LA region of interest candidate refueling/recharging stations

- **Count and Weight Statistics:** Vehicle weight and classification data are sourced from the FAF and the Truck Monitoring and Analysis System (TMAS), both of which aggregate real-world measurements from various monitoring stations [14]. These stations, managed by state highway and transportation agencies, collect essential data on vehicle volume, classification, and weight [14,22]. Whether permanent or temporary, these stations play a vital role in understanding roadway usage. A key technology used is the Weigh-in-Motion (WIM) system, which captures vehicle characteristics, such as weight, axle configurations, and more, as vehicles pass at regular highway speeds. Unlike static scales that require trucks to stop, WIM systems collect dynamic data, including axle loads, spacing, speed, direction, FHWA vehicle classification, and time stamps, without interrupting traffic flow—see Figure 15. Operating continuously, WIM systems provide a rich dataset for analyzing truck volumes and weights, offering critical insights for transportation planning and management. Previous work reported by the authors have demonstrated the potential for reconstructing traffic flows across national highways using limited vehicle classification data from the fixed stations [22]. The OR-AGENT framework leverages this by using an iterative process to impute traffic volume and vehicle class information across broader traffic networks, further enhancing transportation analysis capabilities. Future refinement of this process for weight imputation is currently underway.

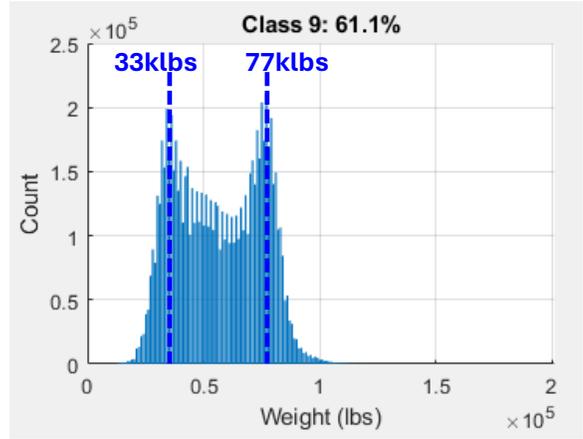

Figure 15. Texas triangle HD tractor-trailer weight distributions (bimodal peaks noted at 33klbs (empty tractor trailer) and 77klbs

C. **Mission Processing:** This process relies on high-resolution vehicle and powertrain models that simulate the key dynamics of vehicles operating in environments described by the ODC. Although the detailed development of these vehicle models falls outside the scope of this paper, it has been extensively covered in prior work [18]. In brief, we have created a 1-D model for heavy-duty diesel, BEV, and FCEV powertrains. The BEV and FCEV architectures are based on a tandem e-axle configuration, with 250 kW electric motors powering both axles. These e-axles feature electric motors integrated into three-speed gearboxes, with additional gear reductions at the axle differential and wheel ends—mirroring state-of-the-art technology for heavy-duty electric trucks. The powertrain models use a forward-looking, quasi-static approach. In prior work published [22], they describe an enhanced driver model that generates realistic speed profiles based on route features like speed limits and road grades, moving away from traditional fixed drive cycles. This more dynamic approach enhances the accuracy of vehicle performance simulations. Additionally, Shiledar et al. developed a road load model that accounts for resistive forces such as aerodynamic drag and tire rolling resistance experienced by trucks in motion. The aerodynamic model adjusts the drag coefficient based on truck configuration (e.g., with or without a trailer) and the yaw angle relative to the wind direction [24]. The tire rolling resistance model accounts for changes in rolling resistance based on vehicle speed and tire temperature, with tire thermal dynamics approximated using a first-order transfer function [25]. The vehicle simulator goes beyond powertrain modeling to include auxiliary components such as cabin heating, ventilation, air conditioning (HVAC) compressors, battery thermal management systems, and pneumatic brake pumps. The energy consumption of these systems is modeled using a duty cycle-based approach, accounting for the influence of ambient conditions on power usage [26,27]. Battery and charger models of similar resolution are also incorporated into the simulator, capturing the electrical, thermal, and aging dynamics of various battery chemistries. This integrated model calculates the energy consumption of trucks over an entire year, factoring in seasonal variations in elevation, road grade, ambient temperature, and air density across the multiple routes identified in the ODC. The comprehensive simulation provides crucial insights into vehicle energy usage in real-world operational environments, enabling more accurate planning for alternative fuel station siting. Battery energy and fuel cell power capacity assumptions will be described in the *Results*.

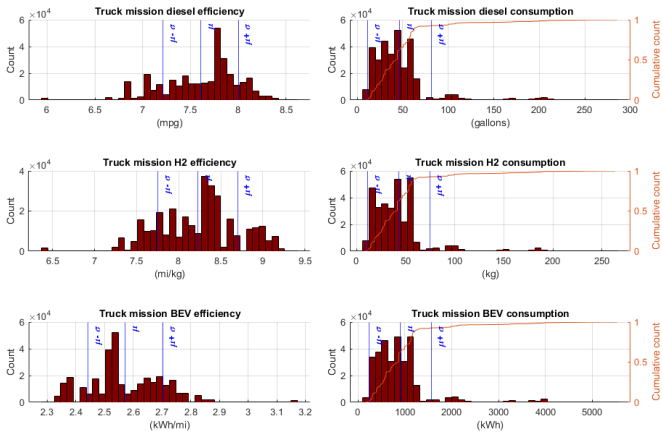

*Figure 16. H2LA region of interest candidate refueling/recharging stations*

D. **Smart Energy Processing:** As outlined in the *Admissible Stops* section, energy processing at both fixed and flexible stopping points can be determined based on the specific mission requirements.

- **Fixed Stopping Points**: Vehicle schedules, trip energy demands, and stop dwell times, combined with the assumption that vehicles begin their day with fully charged batteries or full hydrogen tanks, help calculate the energy required at each stop. The configuration of dispensers or chargers then translates these energy needs into flow or power demand at that stop. By aggregating the energy consumption of all vehicles arriving at a site throughout the year, a clear demand projection can be established. However, the primary challenge lies in the level of detail available for vehicle tours—i.e., not just individual OD pairs but the entire sequence of trips that the vehicle undertakes over the year. If this sequence is known, energy demand can be accurately calculated by tracing the series of ODs. If the sequence is unknown, assumptions must be made regarding the vehicle's energy replenishment needs at each stop. This can be approached in a few different ways. For instance, each trip could be evaluated independently, with the assumption that the vehicle starts with full onboard energy storage and refills to full capacity upon reaching its destination. Alternatively, a sequence of trips could be considered by logically combining factors such as vehicle hours of service, departure schedules across different locations, daily miles traveled, and energy required for the next leg of the trip. This more complex modeling ensures a robust energy demand assessment across multiple operational scenarios. This has been developed in more detail in prior work [15].

- **Flexible Stopping Points:** for refueling or recharging trucks, the process is more dynamic than with fixed stops. Flexible stopping points allow trucks to choose from a list of potential refueling or recharging locations during their trip, rather than being confined to a predetermined set of stations. This flexibility provides opportunities for optimization but also introduces additional complexity into the decision-making process. Each truck tour or trip begins with a clear destination in mind, which represents the last viable stop where the vehicle must arrive with enough energy. During the trip, however, the truck may need to stop at intermediate refueling or recharging stations to ensure it has enough energy to complete its journey. This means that, along each leg of the trip, the vehicle's energy consumption is monitored. If the truck's remaining energy is insufficient to reach the next destination, it must stop at one of the available candidates refueling or charging sites. The need for a stop is determined by evaluating the truck's energy consumption. If the remaining energy falls below a critical level, the model identifies one of the candidate sites as a necessary stop. These candidate locations are pre-identified and distributed along potential routes, allowing flexibility in the truck's path while ensuring sufficient energy levels. Optimization comes into play when determining the total number of stops. This is done by defining an objective function—such as minimizing the number of available stops, minimizing the number of vehicle stops, reducing $CO_2$ emissions, or maximizing energy efficiency. For instance, the model could seek to minimize the total number of refueling or recharging stops needed to complete the trip, or it could prioritize stops that offer renewable energy sources to lower environmental impact. The optimization balances factors like the distance between stops, the truck's energy consumption, and the availability of energy at each station. A failure in this system occurs if the truck is unable to reach its destination due to a lack of suitable refueling or charging stops. This scenario arises when the truck does not stop in time or when no adequate refueling/recharging station is available. Therefore, it is essential to carefully evaluate each trip to ensure that all necessary stops are included along the route to maintain energy levels and avoid operational disruptions. Ultimately, the objective is to optimize the route and the number of stops to ensure that the truck completes its trip efficiently. The optimization process ensures the minimal number of necessary stations are utilized while maintaining energy reserves, supporting sustainability goals, and achieving operational efficiency. This allows the model to recommend the most effective network of refueling and recharging stations along key corridors, balancing operational and environmental factors.

In the H2LA corridor model, the objective function is to minimize the number of available stations in the route network to leave no truck trip stranded (or minimize the number the of stranded trucks), using a flexible set of known stop locations. This is accomplished with a genetic algorithm (GA)—see Figure 17.

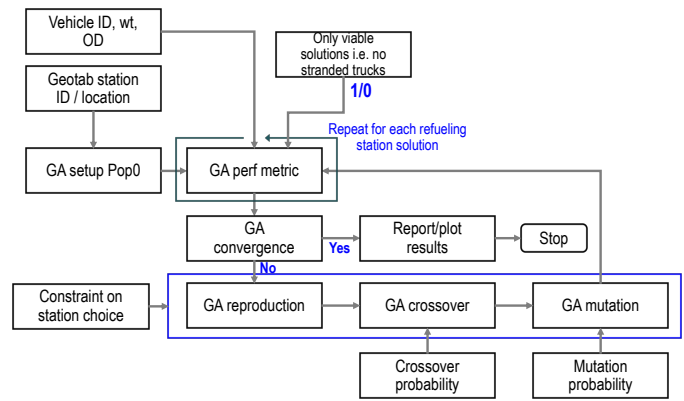

(a) *Overall GA workflow*

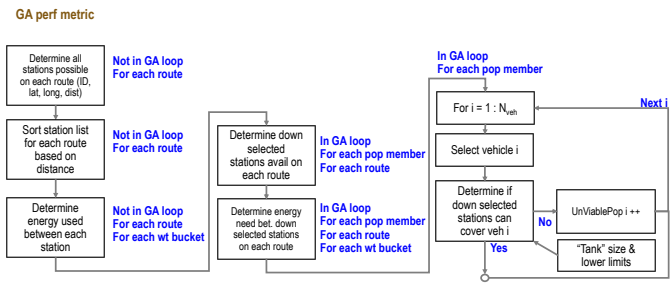

(b) GA performance metric workflow

Figure 17. H2LA smart energy processing optimization using genetic algorithm

- **Smart Charge Management (SCM)** features focus on optimizing the charging process for electric vehicle batteries to enhance efficiency, minimize degradation, and align with grid constraints. At present these are under development for the OR-AGENT model framework (including extensions for real time controls) and will be reported out in future publications..

E.  **Infrastructure Capacity, Cost, Carbon processing**

- **Grid Capacity:** To determine grid capacity at a specific location for a point of use, the following steps are typically undertaken, focusing on evaluating generating capacity, substation capabilities, and proximity of substations within 15 miles of the off-take site. The assessment begins by ensuring all nearby energy generation plants are active and capable of providing power. Next, the available transformer capacity is reviewed to ensure it can service additional loads. Line capacity is then evaluated to confirm whether it can handle the necessary power flow to support these new loads. The analysis also includes reviewing the infrastructure from substations to the end-user sites, such as domiciles or locations where new charging stations will be installed. This step ensures that the power flow can be sustained all the way to the distribution substations, short of the final 480V distribution lines that supply end users. Substation performance data, which is vital for this analysis, is extracted from the North American Energy Resilience Model (NAERM https://www.energy.gov/oe/north-american-energy-resilience-model), developed in part by ORNL for electric grid situational awareness. This comprehensive model enables an in-depth assessment of energy resilience and capacity for infrastructure planning. Using this, a detailed view of grid capacity for each charging location can be developed, ensuring that substations within a 15-mile radius are suitable for providing the required energy for the electric vehicle chargers. Figure 18 shows the regional grid capacity assessment for the H2LA corridor. Due to data sensitivity only aggregate values of substations within a 15-mile radius are shown.

- **Grid Carbon Intensity:** Grid carbon intensity, measured in grams of $CO_2$ per kilowatt-hour (gCO$_2$/kWh), is directly tied to the energy sources supplying the grid. Transitioning from diesel or fossil fuels to electricity does not eliminate a vehicle's carbon footprint but shifts it to the grid, which still relies on carbon-emitting sources such as coal, natural gas, and oil. Even low-carbon sources like nuclear, solar, wind, and hydroelectric power have non-zero carbon intensities due to factors like raw material production and energy transfer losses. By 2021, 40.6% of U.S. electricity was derived from renewable and nuclear sources, up from 35.8% in 2016 [28]. However, carbon intensity varies non-linearly due to seasonal changes, demand fluctuations, and energy exchanges between Balancing Authorities (BAs). To estimate the carbon intensity of added loads, a novel approach uses historical data on grid carbon intensity and demand, tailored to each region [15].

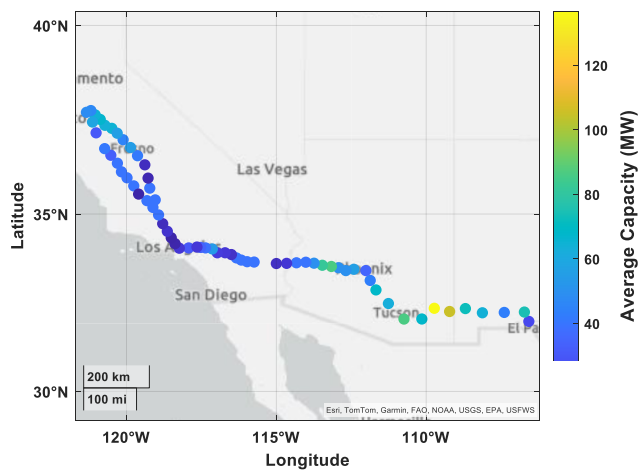

Figure 18. H2LA regional substations and their excess capacity

- **Hydrogen and Natural Gas – Capacity, Cost and Carbon** assessment tool development is currently under exploratory discussions and will be reported out in future publications. This includes using the electric grid information for electrolytic hydrogen generation. Further for the scope of this H2LA corridor paper, the infrastructure capacity, cost and carbon impact is not included but will be included in future publications.

F.  **Microgrid Processing**

The microgrid capabilities will enhance energy resilience, sustainability, and efficiency for the charging infrastructure. Key features include: Renewable Energy Integration, Energy Storage Solutions, Grid Independence and Resilience, Demand Response and Load Management, Seamless Integration with Charging Infrastructure, Support for Electrification Goals.

Overall, microgrid capabilities will create a sustainable, efficient, and resilient energy solution for the charging infrastructure, contributing to the goals of reducing greenhouse gas emissions and supporting the electrification of the trucking industry. To develop an integrated microgrid system utilizing DER requires two key steps. First, the capabilities of the DER assets and their siting must be evaluated and strategically aligned—detailed in [29]. Second, the interaction between the DER assets and grid electricity must be optimized to create the most efficient and reliable energy mix—detailed in [30].

G.  **Output Processing**

Final processing to determine the critical outputs of this process are divided into three efforts: TCO development, Objective optimization, and Roadmap development with details developed in [31].

- **TCO Development:** A comprehensive TCO tool has been developed to evaluate the techno-economic implications of transitioning vehicles and their supporting infrastructure. This innovative tool analyzes various energy transition pathways within the proposed framework, considering both vehicles and infrastructure holistically. As shown in Figure 19, the model is structured into three discrete but interconnected modules—

Vehicle, "Local" Infrastructure, and "Regional" Infrastructure—each encapsulating distinct elements essential to the comprehensive evaluation of decarbonization strategies.

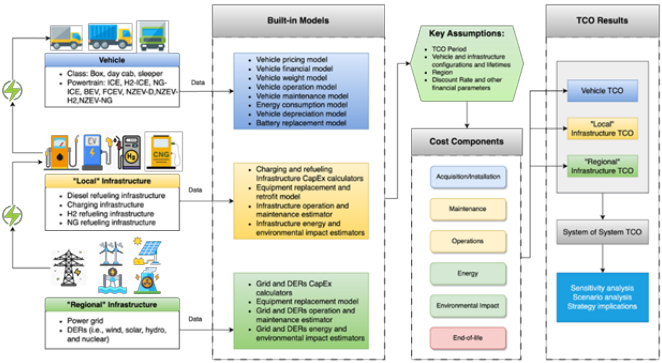

*Figure 19. Overview of the interconnected system TCO tool*

- **Objective Optimization:** The parametric study offers insights by examining specific option studies. If these studies comprehensively cover the problem at hand, the optimal solution (defined by a "cost function") can be easily identified. However, if the search space—combining vehicle types, powertrains, domicile/truck stop configurations, and energy backbone architectures—is too large, a more complex objective optimization process is needed. This process is conducted using a nested structure, as illustrated in Figure 20, and consists of five key steps as shown [15].

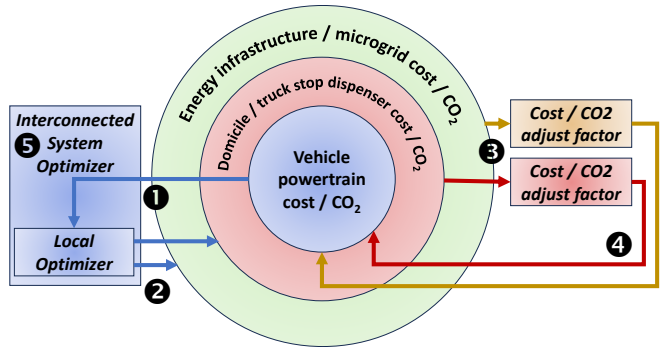

*Figure 20. Simplified nested optimization process (five steps indicated)*

- **Roadmap Development**: After completing the single-year co-optimization—where the fleet of vehicles, powertrain configurations, truck stop architecture, and energy backbone infrastructure are optimized for that year—a roadmap is generated to project how these elements will evolve year by year. This roadmap offers a clear, structured view of how the transportation system will transition over time, considering both technical and economic factors, while satisfying key constraints such as budgetary limitations, carbon emission targets, or ZEV mandates.

## Results and Analysis

### A. Performance Impact Assessment

Using the FAF data to generate OD information, we identified the average annual truck trips and VMT, as illustrated in Figure 21. Based on this data Figure 22 presents the expected well-to-wheel CO2 emissions for various truck propulsion systems assuming a complete conversion of these vehicles from diesel to alternative zero-emission powertrains. Key observations include:

1. Diesel Trucks: The highest level of $CO_2$ emissions comes from diesel trucks, with emissions close to 50 MT/yr. This is expected, given that diesel trucks rely heavily on fossil fuels, leading to higher carbon emissions.

2. BEV: The "BEV 386g/kWh" bar, based on the US electric grid average carbon intensity, shows a significant reduction in $CO_2$ emissions compared to diesel, indicating that electric trucks, even considering emissions from electricity production (likely grid-based), perform better in terms of carbon footprint. However, they still produce $CO_2$, due to grid energy sources still involving fossil fuels.

3. Hydrogen via Steam Methane Reforming ($H_2$ SMR): Hydrogen produced from natural gas (without carbon capture) reduces $CO_2$ emissions compared to diesel but still shows a considerable carbon footprint, due to the reliance on natural gas as the feedstock for hydrogen production.

4. Hydrogen via SMR with Carbon Capture ($H_2$ SMR + CCS): This approach significantly reduces emissions compared to standard SMR, but it is not completely carbon-neutral. The introduction of carbon capture and storage (CCS) technology helps lower the emissions.

5. Hydrogen via Grid Electrolysis ($H_2$ Grid Electrolysis): This method results in the highest $CO_2$ emissions among all hydrogen options, even surpassing diesel, indicating that hydrogen produced via grid-based electrolysis—largely reliant on fossil fuel-generated electricity—can be highly carbon-intensive.

6. Hydrogen via Solar Electrolysis ($H_2$ Solar Electrolysis): This option demonstrates the lowest $CO_2$ emissions, close to zero, showcasing the potential of using renewable energy like solar power to produce hydrogen with minimal environmental impact.

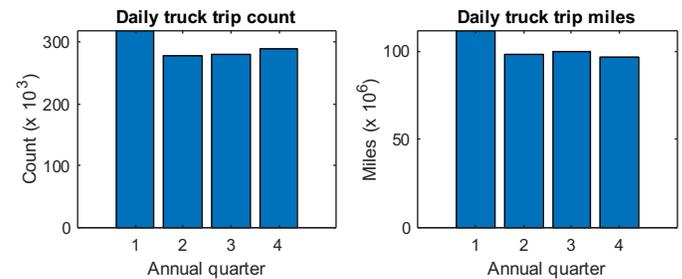

*Figure 21. Daily trips and VMT (using FAF data)*

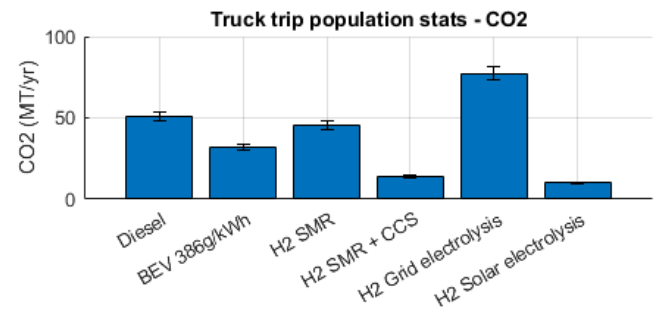

*Figure 22. CO2 emissions impact based on different energy sources*

Figure 23 compares the refueling or recharging rates of various truck propulsion systems in miles of range acquired per minute of refueling or recharging. Key observations include**:**

1. Diesel's Superior Refueling Speed: Diesel trucks refuel at a

significantly faster rate (approximately 275 miles/min) compared to alternative powertrains. This highlights diesel's current dominance in terms of minimizing downtime for refueling, making it attractive for industries prioritizing efficiency and turnaround time.

2. BEV: BEVs with charging capacities ranging from 150 kW to 3750 kW exhibit progressively increasing refueling rates, but even the highest capacity chargers (3750 kW) reach only a fraction of diesel's range gain per minute recharging. The highest-rated BEV charger achieves around 40 miles per minute, showcasing the slower energy transfer rates for electric vehicles and the importance of faster charging solutions for commercial fleets.

3. Hydrogen Refueling Rates: Hydrogen refueling speeds are categorized by the flow rate of hydrogen (in kg/min), with higher range rates as the flow increases. Hydrogen refueling shows rates that can reach close to 100 miles/min at 10 kg/min, though still trailing behind diesel significantly. Nevertheless, hydrogen refueling outperforms electric vehicle charging in terms of miles per minute, especially at higher flow rates.

4. Comparison Between BEVs and Hydrogen: Hydrogen fuel cell vehicles (FCVs) demonstrate faster refueling rates than most BEV charging scenarios, especially at higher hydrogen flow rates. This suggests that hydrogen may present a more competitive refueling solution for long-haul trucks or applications where minimizing downtime is critical.

5. BEV Charging Capacity Matters: The refueling rate increases significantly with the higher charging capacities for BEVs (e.g., from 150 kW to 3750 kW), showing the impact that investment in higher power charging infrastructure can have. However, even at the highest BEV charging rates, diesel and hydrogen (at higher flow rates) maintain a clear advantage.

6. Diesel remains the fastest option, but as the industry shifts toward zero-emission vehicles, hydrogen emerges as a more viable option for long-haul and time-sensitive applications due to its relatively fast refueling rates compared to BEVs. Battery electric vehicles, while offering environmental benefits, still face limitations in refueling speed, which could hinder their adoption for long-distance commercial operations unless ultra-fast charging infrastructure becomes more widespread and efficient.

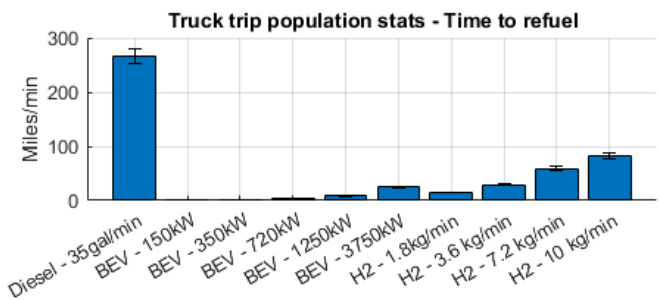

*Figure 23. Refueling time impact based on different energy sources*

In the following analysis for both **hydrogen fueling station siting** and **electric charging station siting**, we assume a 10% conversion of truck trips from diesel to zero-emission powertrains. These results can be scaled based on higher or lower adoption rates. However, certain non-linear effects, such as the number of dispensers or chargers and the corresponding peak power demand, are expected to vary according to the adoption profiles. Additionally, roadmapping for preferred adoption scenarios is illustrated, though not fully developed in this paper, based on the amount of infrastructure deployed. This highlights the importance of strategic planning for both hydrogen and electric vehicle infrastructure to meet future adoption demands.

### B. Hydrogen Fuel Station Siting

Figure 24 illustrates the number of truck trips converted to hydrogen power, highlighting both trips that require refueling and those that can complete the journey without refueling. The optimization metrics previously discussed are applied to ensure that no vehicles are stranded due to fuel shortages, which is confirmed by the data shown in Figure 24. Additionally, we observe that the size of the hydrogen storage tank significantly affects the need for refueling during trips. Larger tanks reduce the frequency of refueling stops, while smaller tanks may require more frequent stops. As a result, the number and location of refueling stations are directly influenced by the vehicle's onboard hydrogen storage capacity, impacting the overall infrastructure planning for hydrogen-powered freight transport.

This optimal solution is achieved by strategically siting hydrogen refueling stations along the extended H2LA corridors, accounting for three different onboard hydrogen storage capacities: 70 kg, 80 kg, and 100 kg. These storage sizes represent typical 700-bar gaseous hydrogen tanks used in heavy-duty trucks, positioned behind the cab. By factoring in varying storage capacities, the stations are optimally located to ensure continuous operation of hydrogen-powered vehicles over long distances, minimizing the need for refueling stops while supporting the decarbonization of freight transport. Figure 25 illustrates the strategic placement of in-route hydrogen refueling stations to support long-distance travel for trucks equipped with these storage capacities. The locations of these stations are designed to align with the key routes shown in Figure 11, Figure 12, and Figure 13, ensuring continuous operation without exceeding the vehicle's fuel range limits. The map highlights critical nodes where hydrogen refueling infrastructure is essential for enabling the feasibility and reliability of hydrogen-powered heavy-duty trucks over vast interstate routes. This infrastructure ensures that hydrogen fuel cell trucks can perform long-haul operations, contributing to the decarbonization of freight transport. By providing refueling stations at pivotal locations, this setup supports the seamless integration of hydrogen technology into the logistics network while reducing emissions and maintaining operational efficiency across the supply chain. In these figures, the size of each circle represents the amount of hydrogen dispensed at each location, with more detailed data provided in Figure 26. Along with the in-route refueling points, each destination replenishes the truck's fuel to full capacity, ensuring readiness for the return trip or subsequent legs. This refueling activity is also depicted in Figure 26. Additionally, it is important to note that we assume each truck begins its trip with a full tank at the origin, and fuel levels are never allowed to drop below 20% capacity to maintain a reserve buffer, ensuring flexibility and avoiding potential fuel shortages. This accounts for the differences in sum of the total fuel dispensed in-route and at the destinations.

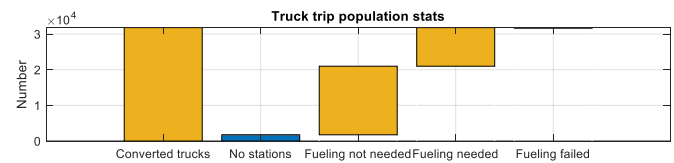

*(a) Hydrogen storage capacity: 70kg*

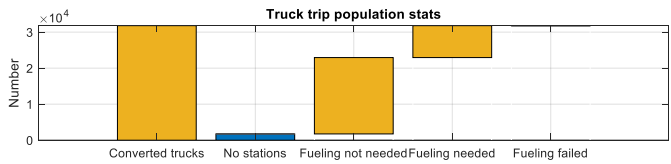

*(b) Hydrogen storage capacity: 80kg*

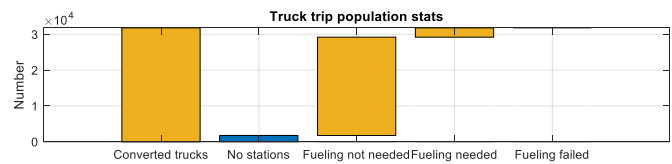

*(c) Hydrogen storage capacity: 100kg*

*Figure 24. Trip population fueling statistics*

This strategic approach underscores the importance of hydrogen as a viable alternative for sustainable long-distance trucking.

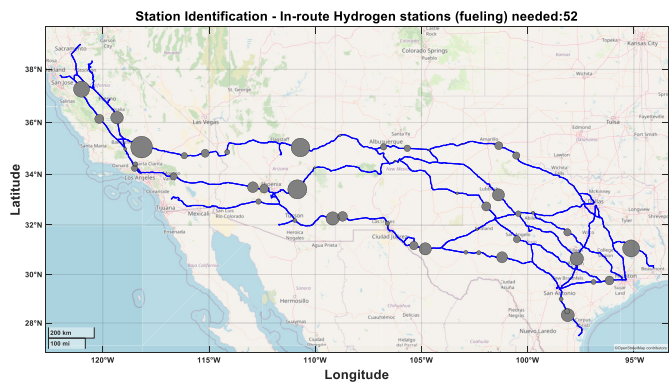

*(a) Hydrogen storage capacity: 70kg*

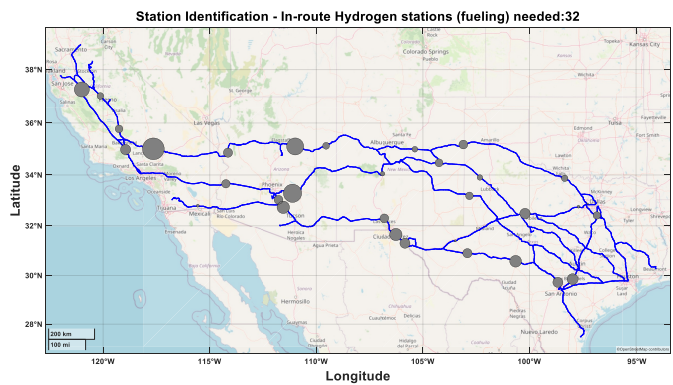

*(b) Hydrogen storage capacity: 80kg*

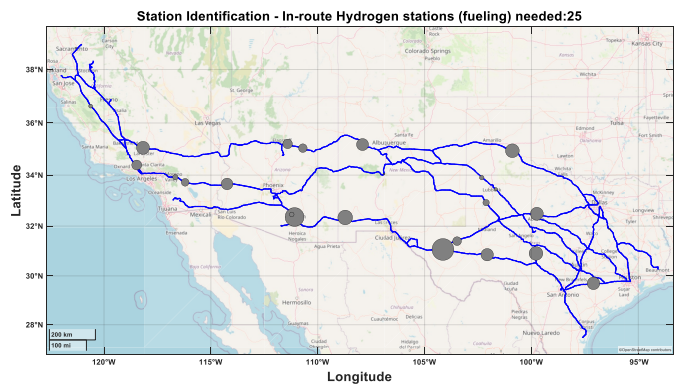

*(c) Hydrogen storage capacity: 100kg*

*Figure 25. Station identification for 100% truck mission completion rate (circle size indicates the amount of hydrogen dispensed)*

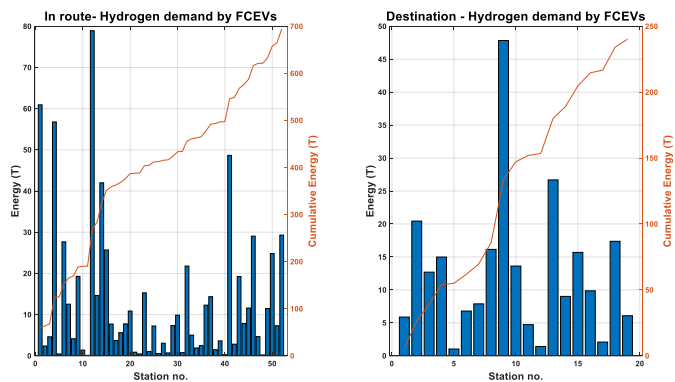

*(a) Hydrogen storage capacity: 70kg*

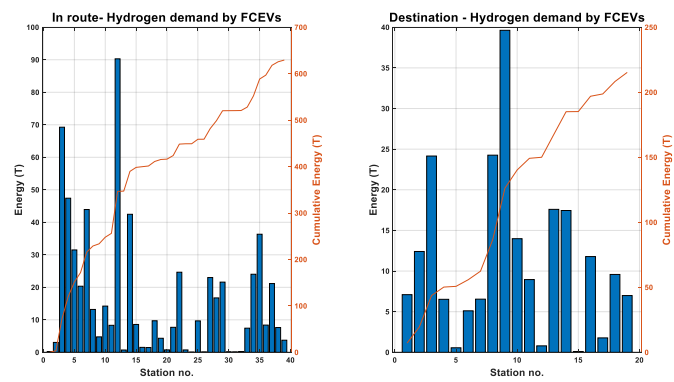

*(b) Hydrogen storage capacity: 80kg*

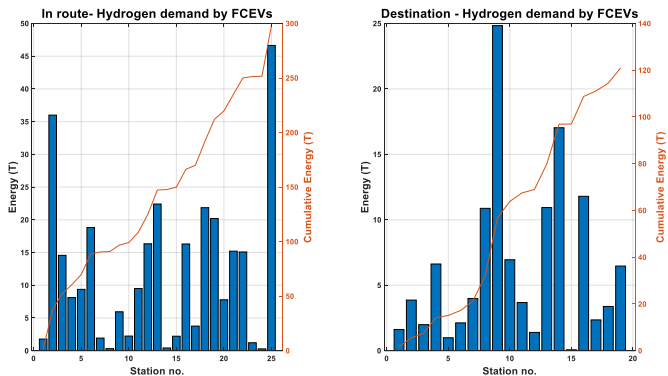

*(c) Hydrogen storage capacity: 100kg*

*Figure 26. Hydrogen dispensed at each site (in-route / destination)*

Figure 27 examines the impact of dispenser types and their associated utilization levels on hydrogen refueling infrastructure. Typical refueling stations aim to maintain an upper utilization limit to ensure dispenser availability when trucks arrive. This target utilization may vary depending on the service provider. For this study, we assume that each station operates with a single dispenser technology (ranging from 1.8 to 10 kg/min) and that all stations in the region target the same utilization level. By setting a uniform utilization target across all stations, we can estimate the number of dispensers of each type required for the region. Figure 27 illustrates this for all in-route refueling stations. As anticipated, larger tanks reduce the need for extensive infrastructure but increase vehicle capital costs. Conversely, smaller tanks demand more infrastructure and require more frequent refueling stops. Since infrastructure costs are typically passed on to the end-user through hydrogen pricing, vehicles with smaller tanks may incur higher operational expenses due to the need for additional refueling infrastructure. This trade-off between tank size, infrastructure requirements, and overall costs underscores the need for further analysis using OR-AGENT to optimize solutions for hydrogen refueling infrastructure development. This ongoing study will help determine the best balance between capital and operational expenses to ensure cost-effective, scalable solutions for the hydrogen trucking ecosystem.

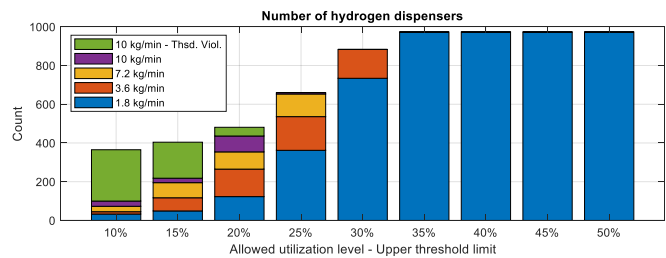

*(a) Hydrogen storage capacity: 70kg*

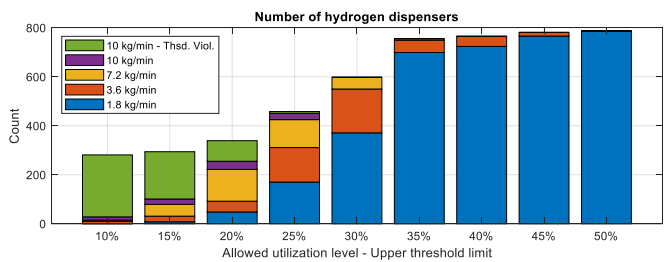

*(b) Hydrogen storage capacity: 80kg*

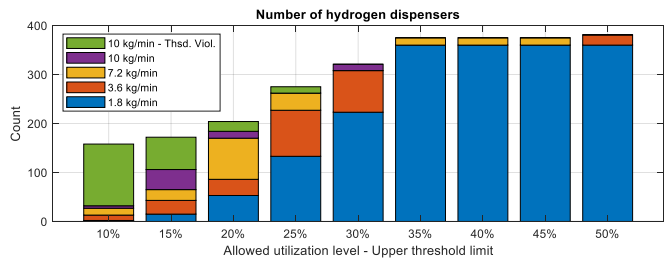

*(c) Hydrogen storage capacity: 100kg*

*Figure 27. Dispenser type count 100% truck mission completion rate*

In Figure 28, we evaluate the impact of reducing the number of hydrogen refueling stations on the percentage of truck trips that become unviable. Initially, 10% of truck trips were randomly selected for conversion to hydrogen powertrains. However, as the number of refueling stations decreases, the percentage of successful trips also declines, with specific trips being eliminated from feasibility rather than random ones. Figure 28 highlights this relationship, showing how a reduced station deployment directly affects the proportion of trips within the initial 10% population that are no longer feasible for hydrogen conversion. This analysis allows for targeted optimization, where, based on a set number of stations, a specific subset of truck trips can be strategically chosen for hydrogen powertrain conversion. This approach ensures that infrastructure limitations are considered in future deployment strategies.

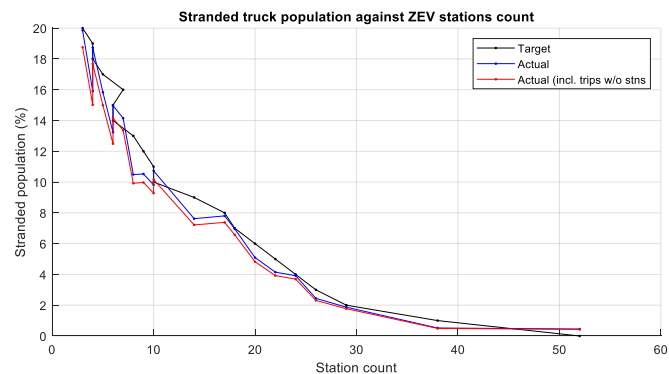

*(a) Hydrogen storage capacity: 70kg*

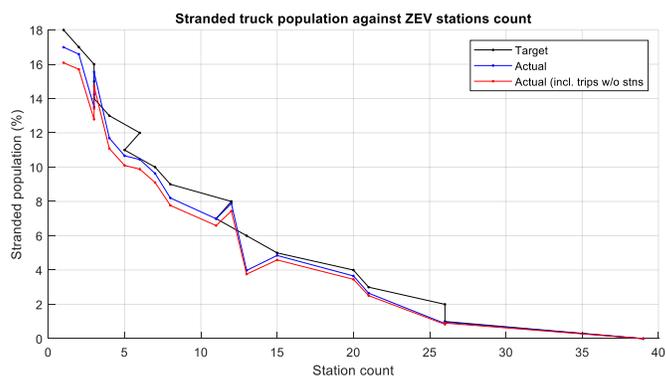

*(b) Hydrogen storage capacity: 80kg*

This manuscript has been authored by UT-Battelle, LLC under Contract No. DE-AC05-00OR22725 with the U.S. Department of Energy. The United States Government retains and the publisher, by accepting the article for publication, acknowledges that the United States Government retains a non-exclusive, paid-up, irrevocable, world-wide license to publish or reproduce the published form of this manuscript, or allow others to do so, for United States Government purposes.

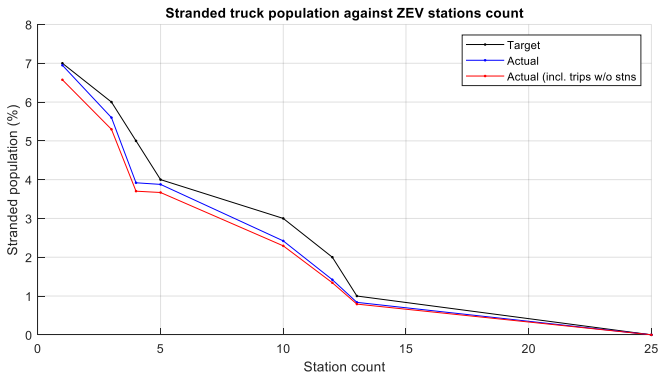

*(c)   Hydrogen storage capacity: 100kg*

*Figure 28. Station quantity based on truck mission completion rate*

### C.  Electric Charging Station Siting

A parallel analysis is conducted for BEVs, focusing on battery capacities ranging from 438 kWh to 1000 kWh, which represent current market technologies such as those seen in the Freightliner eCascadia[4], Nikola Tre BEV[5], and the SuperTruck III program[6]. For this study, the usable battery capacity is set at 80%, reflecting standard operational practices to extend battery life and account for efficiency losses. Additionally, vehicles are programmed to initiate a recharge event when the state of charge (SOC) falls below 15%, creating a buffer to accommodate unforeseen events, such as unexpected delays or detours. This ensures that the vehicle will not face a critical shortage of power during operation. The following figures illustrate the outcomes of this BEV analysis, highlighting the implications of varying battery capacities and state of charge management on vehicle range, charging frequency, and overall operational efficiency. By maintaining these parameters, the study provides insights into the infrastructure needs and logistical considerations for supporting BEV fleets in long-haul operations, while also accounting for the safety and flexibility required in real-world scenarios.

The analysis highlights several critical observations regarding BEVs compared to hydrogen FCEVs in the context of long-haul truck trips. First, as seen in Figure 29, even with the deployment of multiple charging stations, all three battery sizes result in stranded vehicles for part of the population. This indicates that no matter how many stations are selected from the candidate sites, some trucks will not complete their routes without encountering range issues. Additionally, Figure 30 illustrates that BEVs require significantly more charging stations compared to hydrogen refueling stations for the same population of vehicles. While the overall energy dispensed for BEVs may be slightly lower than for hydrogen, as shown in Figure 31, this is offset by the higher efficiency of BEV powertrains. However, a new challenge emerges with the power needed to support BEV charging, given the wide range of charger power levels from 150kW to 1250kW (Figure 32). The substantial infrastructure demands, including the number of chargers and charging spots, further complicate this situation. Lastly, Figure 33 presents the percentage of stranded vehicle trips, clearly showing that even with a high number of charging stations, a significant portion of the population remains stranded. This underscores the complexity of electrifying long-haul freight using BEVs and highlights the critical need for optimized infrastructure to ensure that these vehicles can complete their missions without interruptions.

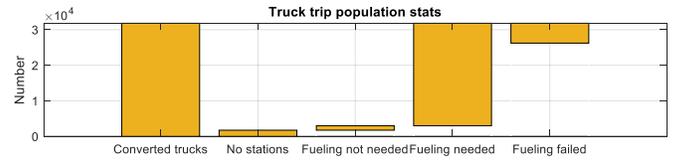

*(a)   Battery storage capacity: 438kWh*

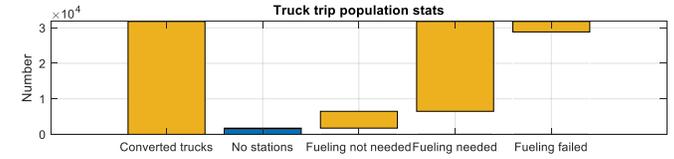

*(b)   Battery storage capacity: 738kWh*

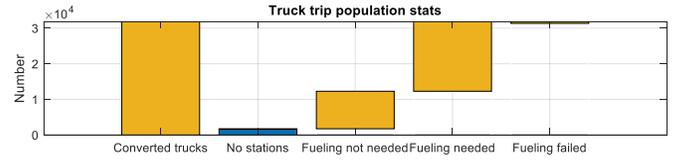

*(c)   Battery storage capacity: 1000kWh*

*Figure 29. BEV truck trip population statistics*

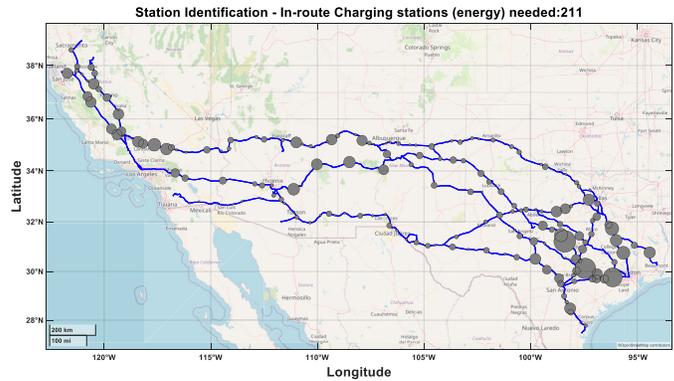

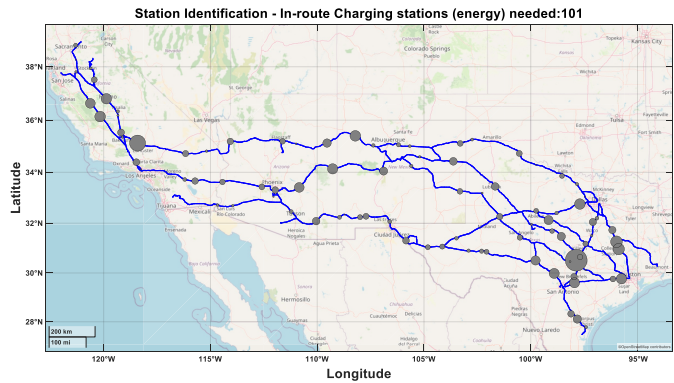

---

[4] Freightliner eCascadia: https://www.freightliner.com/trucks/ecascadia/
[5] Nikola Tre BEV: https://www.nikolamotor.com/tre-bev
[6] SuperTruck III program: https://www.energy.gov/articles/doe-announces-162-million-decarbonize-cars-and-trucks

*(b) Battery storage capacity: 733kWh*

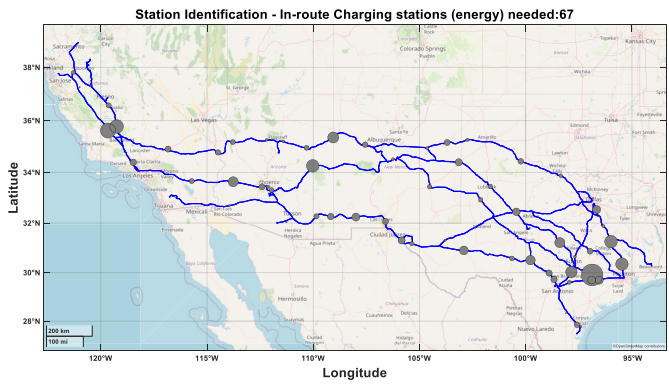

*(c) Battery storage capacity: 1000kWh*

*Figure 30. Station identification for maximum truck mission completion rate (circle size represents the amount of electricity dispensed)*

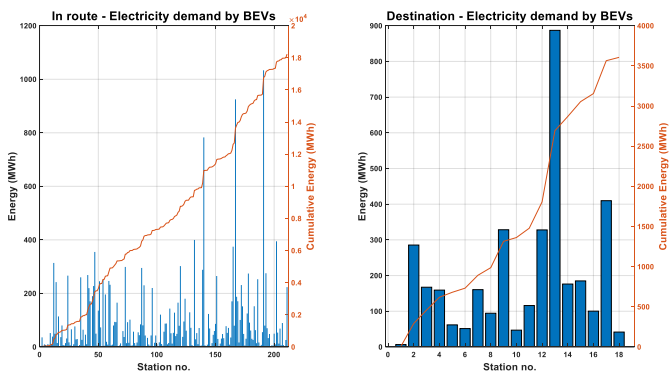

*(a) Battery storage capacity: 438kWh*

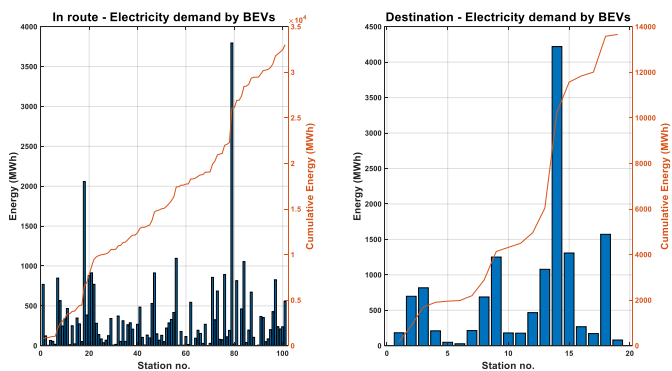

*(b) Battery storage capacity: 733kWh*

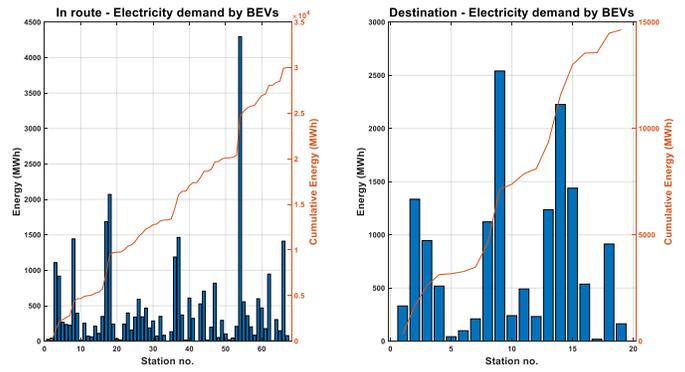

*(c) Battery storage capacity: 1000kWh*

*Figure 31. Electricity dispensed at each site (in-route / destination)*

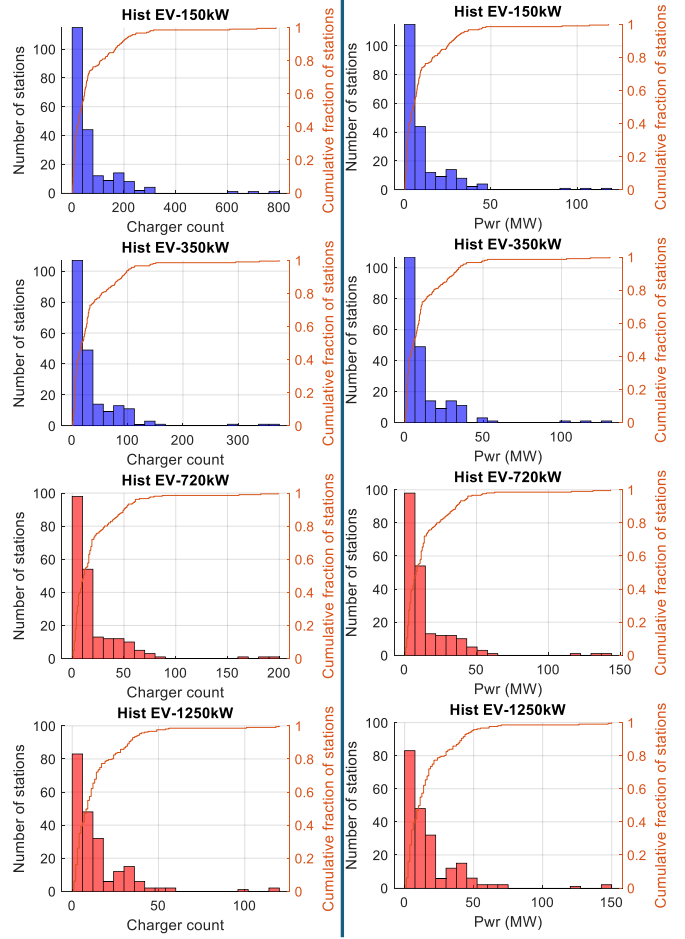

*(a) Battery storage capacity: 438kWh*

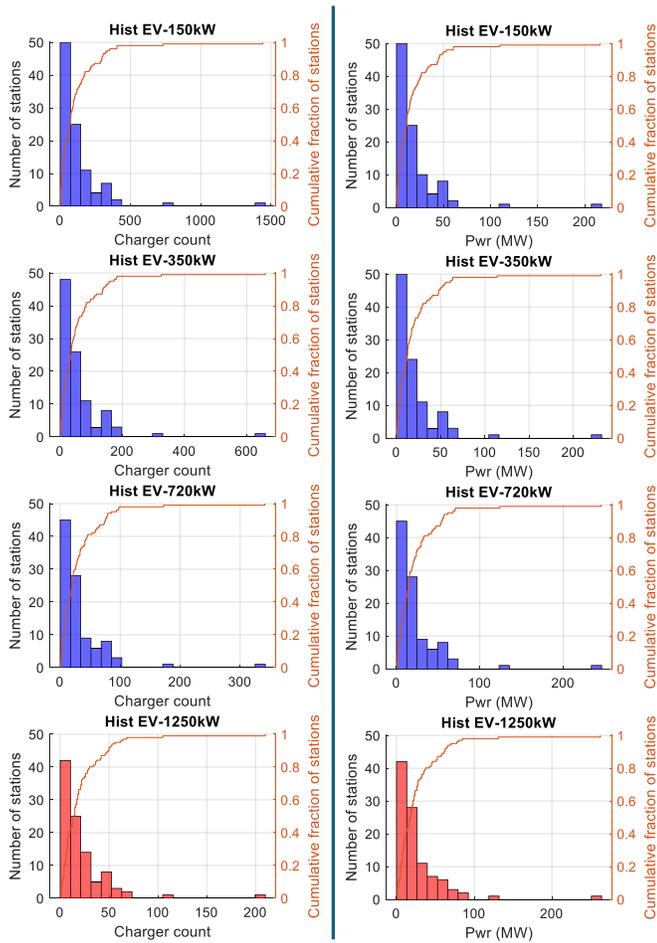

(b)  Battery storage capacity: 733kWh

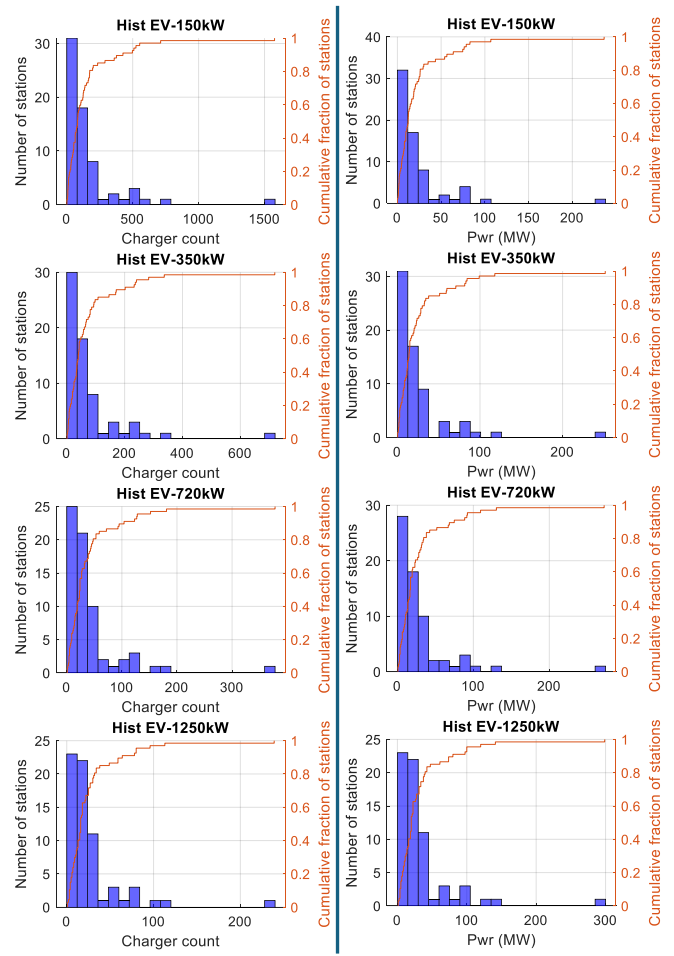

(c)  Battery storage capacity: 1000kWh

Figure 32. Charging power statistics at each site (in-route). Red indicates charging power may exceed 1.5C.

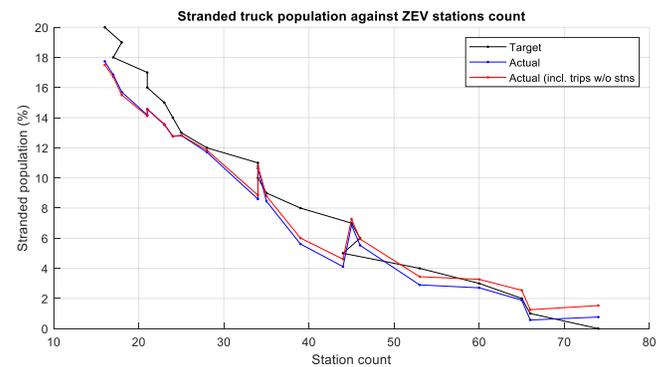

Figure 33. Station quantity based on truck mission completion rate for Battery storage capacity: 1000kWh.

## Discussion

The above analysis shows BEVs and hydrogen FCEVs both offer promising pathways for decarbonizing heavy-duty trucking, but each comes with distinct advantages and challenges. BEVs benefit from higher overall powertrain efficiency and lower operational costs, as electricity is generally cheaper than hydrogen. They also have a more established charging infrastructure, particularly for light and medium-duty vehicles. However, BEVs face significant drawbacks in heavy-duty applications, including long charging times, the need for high-power chargers (150kW to 1250kW), and range limitations, especially over long-haul routes. Additionally, BEV trucks require a larger number of charging stations to avoid "stranded" vehicles, which can be challenging due to the high-power demands and associated infrastructure costs.

In contrast, FCEVs can refuel much faster and offer longer ranges, making them more suitable for long-distance trucking. They also require fewer refueling stations compared to BEVs, reducing the overall infrastructure burden. However, FCEVs face challenges with the current hydrogen production, distribution infrastructure, and higher fuel costs. The carbon intensity of hydrogen production, particularly from fossil fuels, also raises concerns about the overall environmental impact unless green hydrogen production is scaled up. Ultimately, the choice between BEV and FCEV heavy-duty trucks depends on factors like route length, infrastructure availability, and cost considerations.

In addition to establishing siting needs setting up a new multi-megawatt (MW) charging station for electric vehicles (BEVs), or hydrogen refueling stations for fuel cell electric vehicles particularly for commercial fleets and heavy-duty vehicles, requires careful consideration of various technical, operational, and environmental factors. While beyond the scope of analysis of this paper it is noteworthy to highlight these key considerations which include:

A.  **Location and Accessibility**

Proximity to Freight Corridors: The site should be strategically located along major commercial transportation routes or freight corridors to ensure easy access for medium- and heavy-duty vehicles.

Traffic Volume and Demand Forecasting: Anticipated usage levels, based on current and projected traffic volumes of commercial vehicles, should be evaluated to ensure that the station will meet demand without congestion.

Land Availability: Multi-MW charging stations require significant physical space to accommodate multiple high-power chargers, vehicles, and other infrastructure such as transformers and energy storage systems.

Ease of Access: Ensure the station is easily accessible for large vehicles, with ample parking and maneuvering space.

B.  **Energy Supply and Infrastructure**

Grid Capacity: For multi-MW charging stations, the local grid's ability to handle the high electrical loads needs to be assessed. Coordination with utility companies to upgrade grid infrastructure or establish direct connections to substations may be required.

Hydrogen Supply Chain: For hydrogen refueling stations, reliable access to clean hydrogen supply is essential. This may involve proximity to hydrogen production facilities (e.g., electrolysis or steam methane reforming plants), transportation pipelines, or storage facilities.

Energy Storage Systems: Incorporating battery energy storage can help buffer the load on the grid and provide backup power, while hydrogen stations may require on-site hydrogen storage to ensure a constant supply.

Demand Charges: Multi-MW stations are subject to high demand charges from utilities, so energy management strategies (such as energy storage) should be considered to mitigate costs.

Load Balancing: Implement smart charging strategies to distribute power effectively across multiple chargers and minimize peak loads on the grid.

C.  **Refueling/Charging Technology and Capacity**

Charging Speed and Power: Multi-MW stations for commercial vehicles need to provide ultra-fast charging (e.g., 350 kW or higher) to minimize downtime, requiring high-power DC chargers.

Hydrogen Refueling Technology: Stations must be equipped with the appropriate refueling technology, including high-pressure hydrogen dispensers (typically 350 bar or 700 bar) for fast and safe refueling of FCEVs.

Charger/dispensers Configuration: Consider the number of chargers/dispensers and bays required, as well as their layout, to optimize traffic flow and minimize vehicle wait times.

Vehicle Compatibility: Ensure that chargers/dispensers are compatible with the range of vehicles expected to use the station, including connector types and power ratings.

D.  **Site Infrastructure**

Space Requirements: Sufficient space is needed to accommodate large commercial vehicles, turning radii, and parking bays for simultaneous charging or refueling. Additional space is required for hydrogen storage, electrical equipment, and safety buffers.

Utility Access and Upgrades: The site must have easy access to utilities such as water, for cooling and electrolysis (if hydrogen production is on-site), and communications infrastructure for smart grid and monitoring systems.

E.  **Safety and Regulatory Compliance**

Safety Standards: Hydrogen stations must meet stringent safety standards for the storage and handling of hydrogen gas, given its flammability and high pressures. BEV charging stations must comply with electrical safety regulations.

Permitting and Zoning: Securing permits from local, state, and federal authorities is essential. This may include environmental impact assessments, air quality permits, and compliance with local zoning laws.

H2 Safety Protocols: Proper safety measures, such as leak detection, fire suppression, and emergency shutdown systems, are critical for hydrogen stations.

Electrical Safety: Ensure that all electrical components are designed and installed according to safety standards to prevent hazards.

Cybersecurity: Protect the charging infrastructure from cyberattacks, especially when utilizing smart and connected systems.

Physical Security: Secure the site against vandalism or theft, particularly if it's in a remote location.

F.  **Environmental Considerations**

Renewable Energy Integration: If possible, the station should be powered by renewable energy sources (e.g., solar or wind) or offer green hydrogen (produced via electrolysis using renewable energy) to minimize the carbon footprint.

Grid Impact: Assessing and mitigating the potential strain on the electrical grid is important, particularly for multi-MW BEV charging stations. Energy management strategies like demand response may be needed.

G.  **Economic Viability**

This manuscript has been authored by UT-Battelle, LLC under Contract No. DE-AC05-00OR22725 with the U.S. Department of Energy. The United States Government retains and the publisher, by accepting the article for publication, acknowledges that the United States Government retains a non-exclusive, paid-up, irrevocable, world-wide license to publish or reproduce the published form of this manuscript, or allow others to do so, for United States Government purposes.

Capital and Operational Costs: Initial investment costs for infrastructure, grid upgrades, and hydrogen supply must be considered, along with ongoing operational costs such as electricity rates, hydrogen production, and station maintenance.

Revenue and Business Models: A robust business model should account for potential revenue streams from energy sales, public-private partnerships, and incentives from federal or state programs.

H. Scalability and Future-Proofing

Capacity for Expansion: The station should be designed with future growth in mind, allowing for easy expansion to accommodate more charging or refueling points as demand increases.

Technology Advancements: The station should be adaptable to future advancements in charging and hydrogen technology, ensuring it can handle higher power levels or more efficient refueling processes.

I. User Experience and Convenience

Refueling/Charging Speed: For commercial fleets, minimizing downtime is crucial. Charging stations should offer fast charging times, and hydrogen stations should provide rapid refueling capabilities.

Amenities: Stations may need to provide basic amenities for drivers, such as restrooms, food, and overnight parking, particularly for long-haul truck drivers.

Station Monitoring: Use software for real-time monitoring of charger performance, availability, and energy usage to ensure efficient operation and prevent downtime.

Payment Systems: Implement easy-to-use, reliable payment systems that accommodate various forms of payment, such as credit cards, RFID, or mobile apps.

By carefully considering these factors, the setup of a new multi-MW BEV charging station or commercial hydrogen refueling station can effectively support the growing demand for zero-emission commercial transportation while ensuring operational efficiency and long-term sustainability.

## Conclusion

This study underscores the complex but essential task of decarbonizing regional and long-haul freight, given the limitations of BEVs and the infrastructure demands of hydrogen FCEVs. Through a comprehensive analysis using the OR-AGENT framework, the research demonstrates that both BEVs and FCEVs offer viable pathways for decarbonization, but each comes with distinct advantages and challenges.

FCEVs, particularly MHDVs, align well with decarbonization goals set by the Department of Energy and commercial entities. The study highlights that hydrogen-powered trucks have the potential to meet the energy demands of long-haul freight with fewer refueling stations due to their extended range capabilities. However, the carbon intensity of hydrogen production, especially when sourced from fossil fuels, presents a significant challenge. Therefore, the successful deployment of zero-emission hydrogen refueling infrastructure will depend on integrating cleaner production methods, such as renewable or low-carbon hydrogen, to fully realize the environmental benefits. Furthermore, by strategically placing hydrogen refueling stations, the infrastructure can be rolled out affordably and efficiently, with a particular focus on underserved and rural communities that could see improvements in air quality, noise pollution, and energy resiliency.

In contrast, BEVs present a different set of challenges. While they boast higher powertrain efficiency, their limited range and high demand for charging infrastructure—particularly fast chargers—make them less suitable for long-haul freight without significant grid upgrades. The study found that even with a higher density of charging stations, a considerable number of BEV trips were left stranded, which indicates a need for further optimization in charging station deployment and grid capacity. Additionally, the high-power demands for BEV charging infrastructure (due to the concurrent use of multiple chargers ranging from 150kW to 1250kW), exacerbate the strain on the grid, complicating their broader deployment for long-haul operations.

The research primarily focuses on key freight corridors in the Texas Triangle Megaregion (I-45, I-35, and I-10), as well as the I-10 corridor between San Antonio, TX, and Los Angeles, CA, and the I-5/CA-99 corridors in California. These routes are crucial for U.S. freight movement. By using the OR-AGENT framework, the study identifies optimal locations for hydrogen refueling stations and FCEV refueling or BEV charging stations, with the objective of minimizing the number of vehicles stranded along these high-volume freight corridors. Additionally, it offers a first view roadmap for hydrogen refueling station deployment based on different adoption trajectories, aiming for a strategic and scalable rollout.

Future work will focus on:

- Quantified roadmap for hydrogen or BEV station and truck route conversion

- Optimum distribution of combining of both BEV and FCEV technologies which may be necessary to meet the diverse needs of the heavy-duty trucking sector. While BEVs are better suited for shorter, regional hauls due to their efficiency, FCEVs show more promise for long-haul applications, particularly if hydrogen production is decarbonized.

This dual approach can support the transition to a zero-emission freight system, reducing greenhouse gas emissions, improving air quality, and enhancing energy resiliency, especially for rural and energy-stressed communities. Effective infrastructure planning and deployment are critical to overcoming the challenges seen with alternative fuels in the past and ensuring the success of decarbonization efforts in commercial freight transport.

## Acknowledgements

This material is based upon work supported by the U.S. Department of Energy's Office (DOE) of Energy Efficiency and Renewable Energy (EERE) under the Vehicle Technologies Office (VTO) Fiscal Year 2022 Vehicle Technologies Office Program Wide Funding Opportunity Announcement Award Number DE-EE0010650. The authors would also like to acknowledge Ben Gould / Jesse Adams (DOE), Bart Sowa (GTI), Dr. Emily Beagle / Michael Lewis / Lea Daniel, (University of Texas, Austin), Chris Gurciullo / Cate Kehn (ExxonMobil), Hank Murphy / Marco Ugarte / Ben Whitmore (Walmart) for their ongoing support of this research project.

## Contact Information

Vivek Sujan, sujanva@ornl.gov

## Definitions/Abbreviations

| | |
|---|---|
| **BA** | **Balancing Authorities** |
| **BEV** | **Battery Electric Vehicle** |
| **CH4** | **Methane** |
| **CO2** | **Carbon Dioxide** |
| **CV** | **Commercial Vehicles** |
| **DER** | **Distributed Energy Resources** |
| **DOE** | **Department of Energy** |
| **DOT** | **Department of Transportation** |
| **EEJ** | **Environmental Equity and Justice** |

| | | | |
|---|---|---|---|
| EIA | Energy Information Administration | V2G | Vehicle to Grid |
| EPA | Environment Protection Agency | V2H | Vehicle to Home |
| FAF | Freight Analysis Framework | VMT | Vehicle Miles Traveled |
| FCEV | Fuel Cell Electric Vehicle | WIM | Weigh in Motion |
| FHWA | Federal Highway Administration | WIND | Wind Integration National Dataset |
| GA | Genetic Algorithm | ZEV | Zero Emission Vehicle |
| GHG | Greenhouse Gas | | |
| GIS | Geographic Information Systems | | |
| GPS | Global Positioning System | | |
| H2Hubs | Hydrogen Hubs | | |
| H2LA | Houston to Los Angeles | | |
| HVAC | Heating, Ventilation, Air conditioning | | |
| ICE | Internal Combustion Engine | | |
| IIJA | Infrastructure Investment and Jobs Act | | |
| Kg | Kilogram | | |
| kW | Kilowatt | | |
| kWh | Kilowatt-hour | | |
| LCOE | Levelized Cost of Energy | | |
| MHDV | Medium and Heavy-Duty Vehicles | | |
| MILP | Mixed Integer Linear Programming | | |
| MW | Megawatt | | |
| N2O | Nitrous Oxide | | |
| NAERM | North American Energy Resilience Model | | |
| NEVI | National Electric Vehicle Infrastructure | | |
| NHTS | National Household Travel Survey | | |
| NOAA | National Oceanic and Atmospheric Administration | | |
| NSRDB | National Solar Radiation Database | | |
| OD | Origin-Destination | | |
| ODC | Operating Domain Characterization | | |
| ODS | Operating Domain Specification | | |
| OR-AGENT | Optimal Regional Architecture Generation for Electrified National Transport | | |
| OR-SAGE | Oak Ridge Siting Analysis for Power Generation | | |
| ORNL | Oak Ridge National Laboratory | | |
| reV | Renewable Energy Potential | | |
| SCM | Smart Charge Management | | |
| TCO | Total Cost of Ownership | | |
| TEU | Twenty-foot Equivalent Unit | | |
| UTC | Coordinated Universal Time | | |